\documentclass{jfm}
\usepackage{float}
\usepackage{graphicx}
\usepackage{epstopdf, epsfig}
\usepackage{gensymb}
\usepackage[dvipsnames]{xcolor}
\usepackage{pgf,tikz}

\usepackage{tikz, pgf, pgfplots, wrapfig}
\pgfplotsset{compat=newest}

\usetikzlibrary{automata,positioning,arrows,backgrounds,plotmarks, external, 
	calc}
\usepackage{pgfplots}
\usepackage{pgfplotstable}
\usetikzlibrary{calc}
\usepgfplotslibrary{groupplots}

\usetikzlibrary{plotmarks}
\usetikzlibrary{arrows.meta}
\usepgfplotslibrary{patchplots}
\usepackage{grffile}

\usepackage{colortbl} 
\usepackage{multirow} 

\usetikzlibrary{backgrounds}

\usepgfplotslibrary{external} 
\usetikzlibrary{external}             
\pgfplotsset{compat=newest,table/search path={figs}} 

\usepackage{amsmath}
\usepackage{amssymb}
\usepackage{color,soul}
\usepackage{physics}

\definecolor{mycolor1}{rgb}{0.00000,0.44700,0.74100}%
\definecolor{mycolor2}{rgb}{0.85000,0.32500,0.09800}%
\definecolor{mycolor3}{rgb}{0.92900,0.69400,0.12500}%
\definecolor{mycolor4}{rgb}{0.49400,0.18400,0.55600}%
\definecolor{mycolor5}{rgb}{0.46600,0.67400,0.18800}%

\newcommand*{\solidline}{
	{\color[rgb]{0.00000,0.44700,0.74100}
		\begin{tikzpicture}[baseline=-0.5ex]

		\draw [line width=1.0pt] (0,0) -- (5ex,0);

		\end{tikzpicture}
}}

\newcommand*{\dashedline}{
	{\color[rgb]{0.85000,0.32500,0.09800}
		\begin{tikzpicture}[baseline=-0.5ex]

		\draw [dashed,line width=1.0pt](0,0) -- (5ex,0);

		\end{tikzpicture}
}}

\newcommand*{\dottedline}{
	{\color[rgb]{0.92900,0.69400,0.12500}
		\begin{tikzpicture}[baseline=-0.5ex]

		\draw [dotted,line width=1.0pt](0,0) -- (5ex,0);

		\end{tikzpicture}
}}

\newcommand*{\dashdotted}{
	{\color[rgb]{0,0,0}
		\begin{tikzpicture}[baseline=-0.5ex]

		\draw [dashdotted,line width=0.1pt](0,0) -- (5ex,0);

		\end{tikzpicture}
}}

\newcommand*{\orangesquare}{
	{\color[rgb]{0.85000,0.32500,0.09800}
		\begin{tikzpicture}

		\fill (-0.05,-0.05) rectangle (0.1cm,0.1cm);

		\end{tikzpicture}
}}

\newcommand*{\bluecircle}{
	{\color[rgb]{0.30,0.30,1.00}
		\begin{tikzpicture}

		\fill (2,2) circle (0.08cm);

		\end{tikzpicture}
}}

\usepackage{relsize}

\usepackage{subcaption}
\usepackage{animate}
\DeclareMathOperator{\diag}{diag}
\newcommand\fft{{${m-{\omega}}$ }}
\newcommand{\hermconj}{^{\mathsf{H}}}
\usepackage{animate}

\makeatletter
\newcommand\thefontsize[1]{{#1 The current font size is: \f@size pt\par}}
\makeatother

\usepackage{blindtext}
\usepackage{textcomp}
\usepackage{hyperref}
\hypersetup{%
	bookmarksnumbered,
	pdfstartview=FitH,
	unicode,
	pdfborder = {0 0 0},
	colorlinks,
	citecolor=blue,
	linkcolor=blue,
}
\usepackage{url}
\usepackage{lipsum}  

\shorttitle{Nozzle dynamics and wavepackets in turbulent jets}
\shortauthor{O. Kaplan, P. Jordan, A. V. G. Cavalieri and G. A. Br\`{e}s}
\title{Nozzle dynamics and wavepackets in turbulent jets}

\author{O{\u g}uzhan  Kaplan\aff{1,4}
  \corresp{\email{oguzhan.kaplan@centralelille.fr}},
  Peter Jordan\aff{1},
  Andr{\'e} V. G. Cavalieri\aff{2}
 \and {Guillaume A. Br\`{e}s} \aff{3}}

\affiliation{\aff{1}Institut Pprime, CNRS - Universit\'{e} de Poitiers - ENSMA, Poitiers, France
\aff{2}Divis\~{a}o de Engenharia Aeronautica, Instituto Tecnologico de Aeronautica, 12228-900 S\~{a}o Jose dos Campos, SP, Brazil
\aff{3}Cascade Technologies Inc., Palo Alto, CA, USA
\aff{4} Univ. Lille, CNRS, ONERA, Arts et Metiers ParisTech, Centrale Lille, FRE 2017 - Laboratoire de M\'{e}canique des fluides de Lille - Kamp\'{e} de Feriet, Lille, France}

\begin{document}

\maketitle

\begin{abstract}

\noindent We study a turbulent jet issuing from a cylindrical nozzle to characterise coherent structures evolving in the turbulent boundary layer. The analysis is performed using data from a large-eddy simulation of a Mach 0.4 jet. Azimuthal decomposition of the velocity field in the nozzle shows that turbulent kinetic energy predominantly resides in high azimuthal wavenumbers; the first three azimuthal wavenumbers, that are important for sound generation, contain much lower, but non-zero amplitudes. Using two-point statistics, low azimuthal modes in the nozzle boundary layer are shown to exhibit significant correlations with modes of same order in the free-jet region. Spectral Proper Orthogonal Decomposition (SPOD) is used to distill a low-rank approximation of the flow dynamics. This reveals the existence of tilted coherent structures within the nozzle boundary layer and shows that these are coupled with wavepackets in the jet. The educed nozzle boundary-layer structures are modelled using a local linear stability analysis of the nozzle mean flow. Projection of the leading SPOD modes on the stability eigenmodes shows that the organised boundary-layer structures can be modelled using a small number of stable eigenmodes of the boundary-layer branch of the eigenspectrum, indicating the prevalence of non-modal effects. Finally local and global resolvent analysis of the mean-flow are performed. It is shown that the most-energetic nozzle structures can be successfully described with optimal resolvent response modes, whose associated forcing modes are observed to tilt against the nozzle boundary-layer, suggesting that the Orr mechanism underpins these organised, turbulent, boundary-layer structures.

\end{abstract}


\section{Introduction}

Azimuthally organised coherent structures in the form of wavepackets are shown to play a central role in turbulent jet dynamics where sound generation is concerned \citep{mollo1967}. These structures have correlation lengths greater than the integral scales of the turbulence field in which they evolve. Flow visualisations supporting their existence have been reported in studies of forced and natural jets by \citet{crow1971orderly} and \citet{moore1977role}. Since the first efforts by \citet{crow1971orderly}, \citet{michalke1971instabilitaet}, and \citet{crighton1976stability}, wavepacket modelling has relied on some form of the linearised Navier-Stokes equations. Wavepackets are thus interpreted as small amplitude perturbations that evolve on the turbulent mean flow, which acts as an `equivalent laminar base flow' \citep{crighton1976stability}. Recent reviews have been provided by \citet{jordan2013wave} and \citet{cavalieri2019wave}.

In subsonic turbulent jets, wavepackets undergo spatial growth in the initial region downstream of the jet-exit. Studies by \citet{suzuki2006instability}, \citet{gudmundsson2011instability} and \citet{cavalieri2013wavepackets} used mean-flow based linear analysis to describe near-field pressure and velocity fluctuations associated with wavepackets; their results illustrate that Kelvin-Helmholtz instability is the main driving mechanism for wavepacket growth until the point at which the wave becomes naturally stable \citep{crow1971orderly}. Further downstream, various studies show that non-modal mechanisms play an important role for wavepacket dynamics \citep{jordan2017modal,tissot2017wave,lesshafft2019resolvent, schmidt2018spectral}. These non-modal effects are shown to stem from non-linearities that may be associated with non-linear wave interactions, or background turbulence. The recent emergence of resolvent analysis of turbulent flows provides a means by which to link the non-linear and linear dynamics that underpin wavepackets in jets; the non-linear terms are interpreted as an external forcing that drives the linear flow response \citep{sharma2013coherent}. Applications in wall-bounded flows, flow over backward-facing steps, and more recently in turbulent flames are reported in the works of \citet{mckeon2010critical}, \citet{beneddine2016conditions} and \citet{kaiser2019prediction}.

Spectral proper orthogonal decomposition (SPOD), first proposed by \citet{lumley1967structure}, proves to be an effective tool for educing coherent structures in turbulent flows, and has been employed frequently in the analysis of numerical and experimental data of turbulent jets and wall-bounded flows \citep{citriniti2000reconstruction,towne2018spectral,lesshafft2019resolvent,muralidhar2019spatio,abreu2020spod}. Furthermore, recent works by \citet{semeraro2016modeling}, \citet{towne2018spectral} and \cite{lesshafft2019resolvent} have drawn a parallel between resolvent and SPOD analyses: resolvent response modes and SPOD modes are equivalent if the non-linear forcing field is spatially white. However, it is clear that the forcing in turbulent flows is not white noise in space, and recent works in wall-bounded turbulence have shown that non-linear terms are quite structured \citep{nogueira2020forcing,morra2020colour}. Nonetheless, if the resolvent operator has significant gain separation, the leading response mode from resolvent analysis is expected to be close to the first SPOD mode regardless of the specific details of the forcing \citep{beneddine2016conditions,cavalieri2019wave}. Such observations identify SPOD as an ideal signal-processing tool for use in parallel with resolvent analysis.

For Strouhal numbers---based on jet-exit values--- above $St>0.3$ and a range of azimuthal modes, wavepacket development appears to undergo a two-stage process \citep{schmidt2018spectral,lesshafft2019resolvent}: initial wavepacket evolution can be characterised using an optimal resolvent response mode with an associated forcing mode, either localised near \citep{schmidt2018spectral} or distributed within \citep{lesshafft2019resolvent} the nozzle, and with virtually zero support in the downstream region. This leads to spatially amplifying disturbances due to the Kelvin-Helmholtz (KH) instability mechanism. Beyond the point where the turbulent mean-flow becomes locally, neutrally stable, non-modal effects become important and wavepackets may undergo a spatial non-modal growth underpinned by the Orr mechanism \citep{tissot2017wave,tissot2017sensitivity}, which is activated/excited by non-linear flow interactions that can be modelled as a volumetric forcing of the linear operator in the resolvent framework. The same non-modal mechanism is also shown to underlie wavepacket development at low frequencies in the initial shear-layer region of the jet, which the modal stability models fail to characterise \citep{breakey2017experimental,cavalieri2013wavepackets}.

\subsection{Nozzle-exit conditions and their effect on sound generation}\label{intro_nozzleexit}
Many studies have explored the effect of (initial) nozzle-exit conditions on flow development in jets \citep{bradshaw1966effect,batt1975layer,
hussain1978effects,hussain1978effects_mom,hill1976effects}. These studies have informed and guided research efforts to explore the sensitivity of jet-noise radiation to the initial conditions. The state of the nozzle-exit boundary layer is documented to impact sound generation in subsonic jets \citep{mollo1967,maestrello1971acoustic,grosche1974distributions}. 

Jets with initially laminar boundary layers are shown to emit greater far-field noise than those with disturbed laminar boundary layers with high turbulence intensities ($u^{\prime}/{U_j} \approx 0.1$)   \citep{zaman1985effect,zaman2012effect,viswanathan2004effect}. The augmented noise is shown to be associated with vortex roll-up and pairing dynamics in the initial region of jet, observed experimentally by \citet{bridges1987roles,zaman1981turbulence} and in numerical simulations \citep{bogey2010influence}. The observed differences are particularly pronounced at higher frequencies ($St>1$).

\citet{fontaine2015very} studied the effect of turbulent boundary layers with different thickness but with similar turbulence intensities, and showed that a thicker boundary layer resulted in up to 3 dB reduced far-field sound at low polar angles. The authors assert that because the thicker shear-layer has a missing `initial thickness' at the nozzle exit, the potential core closes earlier, leading to a more constricted volume for sound generation. This results in noise reduction at high frequencies; the associated sound being found to scale with the momentum thickness of the exit shear layer.

Drawing on a wide range of numerical and experimental studies cited above, \citet{bres2018importance} recently explored the relevance of nozzle-exit conditions for jet-noise radiation using a large-eddy-simulation of an isothermal $M=0.9$ jet, and provided a meticulous account of the relation between jet noise and boundary layers at the nozzle exit. The authors conducted a parametric study to characterise the effects of wall modelling, grid refinement near the nozzle interior surface and implementation of synthetic turbulence on the downstream flow development and far-field noise. Comparisons with a companion experiment with turbulent nozzle-exit boundary layer \citep{cavalieri2013wavepackets} emphasised the importance of application of the three methods for predicting flow development and sound radiation. Consistent with the earlier findings in the literature, initially laminar jets are found to emit more noise at high frequencies. Results of the linear stability analysis of the near-nozzle mean flow for the low azimuthal modes, which are shown to be important for sound generation, suggest that this difference is associated with greater amplification of Kelvin-Helmholtz mode near the nozzle exit at high frequencies for laminar jets.

Over the past decades, significant insights are obtained on wavepackets and their importance for sound radiation in turbulent jets. Their signatures, both in the near and far fields, are observed; and the associated dynamics can be described using linear models. But there remain many questions. And among these, what fixes the amplitudes of wavepackets? Non-linear turbulent interactions in the region downstream of the jet-exit and/or upstream conditions? Nozzle-exit boundary-layer conditions are shown to influence both noise generation and jet turbulence. However, it is not clear what role upstream conditions play in the evolution of downstream wavepackets. This raises a question regarding the eventual presence of wavepackets in the upstream region of the nozzle-exit, the relationship between these and downstream wavepackets, and the modelling framework that would describe such structures within the nozzle.

With these questions in mind, we explore the dynamics of a turbulent nozzle boundary layer of a $M=0.4$ jet and broadly address three research questions using a high-fidelity large-eddy simulation database: (1) 
{Characterisation of the dynamics of the nozzle boundary layer for low azimuthal wavenumbers.} (2) {The connection between the associated flow-structures in the nozzle and wavepackets in the downstream region of the jet.} (3) {Modelling these structures using the linearised Navier-Stokes equations.} 

The paper is organised around four sections. \S \ref{les_database} introduces the large-eddy-simulation. We characterise the mean nozzle boundary-layer properties and its azimuthal structure, and move on to explore how low-order nozzle dynamics are connected to wavepackets in the downstream region of the jet in \S \ref{nozzle_flow}. The next two sections are devoted to the characterisation and modelling of flow structures identified in the preceding section. Motivated by the success of linear theory in modelling wavepackets downstream of the jet exit, \S \ref{lst_nozzle} attempts to model low-order azimuthal nozzle boundary-layer structures using a local spatial linear stability analysis of turbulent nozzle mean-flow. Finally in \S \ref{resolvent_nozzle}, an input-output or resolvent analysis is performed. Conclusions and future avenues for research close the paper.

\section{Numerical database : Large-eddy-simulation of a M=0.4 turbulent jet}\label{les_database}

\begin{table}
	\begin{center}
		\begin{tabular}{lccccccc}

			& $x$ & $r$ & ${\theta}$  & $n_x$  & $n_r$  & $n_{\theta}$ \\[10pt]
			
			Jet & $[0,30]$ & $ [0,6]$ & $[0, 2\upi]$ & $656$  &    $138$  & $128$ \\[10pt]
			Nozzle & $[-2.8,0]$ & $ [0,0.5]$ & $[0, 2\upi]$ & $225$  &    $58$  & $128$ \\[10pt]

		\end{tabular}
		
		\caption{Grid parameters of the large-eddy-simulation of the $M=0.4$ jet}\label{t:grid}
		
	\end{center}
\end{table}

The numerical database used is provided by a high-fidelity large-eddy-simulation (LES) of a $M=U_j/{c_{\infty}}=0.4$ isothermal axisymmetric turbulent jet, using the compressible flow solver `Charles' developed at Cascade Technologies \citep{bres2017unstructured,bres2019modelling}. This database is an extension of the previous work by \citet{bres2018importance} and validated for the same nozzle configuration at $M=0.9$. The characteristic scales used in the non-dimensionalisation of velocity, length and time are jet-exit velocity $U_j$, jet diameter $D$ and $t=D/U_j$ respectively. The Reynolds number of the jet with the selected scales is $Re={U_{j}D}/{\nu} \simeq$ 450 000. The cylindrical coordinate system is centered at the jet-exit in the computational domain which extends over $-10<x<50$ in the axial direction and spreads from $r=20$ to $r=50$ in the radial direction in the jet.

Wall modelling and near-wall adaptive mesh refinement are employed on the internal nozzle surface, in addition to synthetic turbulence boundary conditions over $-2.8<x<-2.5$ to mimic the boundary-layer trip in the experiment. The near-wall resolution was chosen based on an initial estimate of the nozzle-exit boundary-layer thickness in the companion experiment by \citet{cavalieri2013wavepackets} to produce approximately 10-20 LES cells in the developing boundary-layer region. The resulting wall-normal grid spacing in wall units $y^{+}_{LES}$ is of order \textit{O}(70), and of order \textit{O}(100) for the streamwise and azimuthal grid spacings for the present case. Therefore, the physics of the viscous sublayer is described using wall modelling, and the results in the (very) near-wall region should be considered with caution. Nevertheless, main boundary-layer dynamics, and in particular structures as large as the boundary-layer thickness are well resolved and accurately captured in the LES, as comparison with experimental measurements will show in a later stage of the paper. The total simulation time in acoustic units is ${t}={t{c_{\infty}}/D}=2000$ and the flow field is sampled with time intervals of ${\Delta t}={\Delta t{c_{\infty}}/D}=0.2$, where  ${c_{\infty}}$ denotes the free-stream speed of sound. The corresponding sampling Strouhal number $St_{s}$ based on jet-exit velocity and diameter is $St_{s}=f_{s}D/{U_j}=12.5$ ($f_{s}=1/{\Delta t}$), providing 10000 snapshots of the flow. The flow is simulated on an unstructured mesh; the LES data is interpolated onto a structured cylindrical grid and its parameters are summarised in table \ref{t:grid}. A comprehensive description of the numerical simulation and the nozzle interior flow modelling can be found in \citet{bres2017unstructured,bres2018importance}.

\section{Analysis of nozzle flow}\label{nozzle_flow}

\subsection{Nozzle mean-flow properties and single-point statistics}

\begin{figure}
	\centering
	\includegraphics[scale=1]{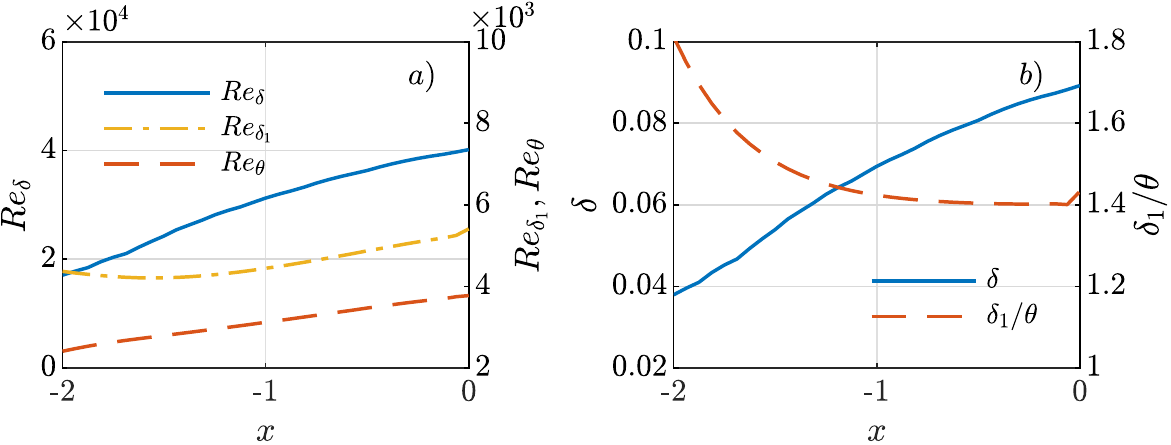}
	\caption{Streamwise evolution of (a) Reynolds number based on nozzle boundary-layer outer units and (b) nozzle boundary-layer thickness and shape factor $H={{\delta}_{1}}/{\theta}$ for $M=0.4$ jet.}
	\label{fig:BLparametersM04}
\end{figure}

The objective of this section is to characterise the mean-flow state throughout the nozzle. Displacement $\delta_1$ and momentum $\theta$ thicknesses of the nozzle boundary layer are estimated as

\begin{equation}
\label{e:disp_thickness}
{\delta_{1}}(x)=\int_{0.5}^{0}  \Bigg (1-\frac{\overline{u_{x}}(x,r)}{\overline{u_{x}}(x,0)} \Bigg) dr. 
\end{equation}

\begin{equation}
\label{e:mom_thickness}
{\theta}(x)=\int_{0.5}^{0} {\frac{\overline{u_{x}}(x,r)}{\overline{u_{x}}(x,0)}} \Bigg (1-\frac{\overline{u_{x}}(x,r)}{\overline{u_{x}}(x,0)} \Bigg) dr,
\end{equation}

\noindent where $\overline{(.)}$ refers to time- and azimuthal-averaged values; $u_x$ denotes streamwise velocity. Figure \ref{fig:BLparametersM04}a shows the streamwise evolution of Reynolds number in terms of boundary-layer units. Nozzle-exit values of the boundary-layer parameters are summarised in table \ref{t:bl_integral}; these are compared with estimations made from hot-wire measurements in the corresponding experiment \citep{cavalieri2013wavepackets}, revealing a good agreement and suggesting that the boundary layer is turbulent at the nozzle exit, which is also corroborated by the value of shape factor $H=\delta_{1}/\theta$ approaching a value typical
for zero-pressure-gradient boundary layers in figure \ref{fig:BLparametersM04}b \citep[p. 423]{schlichting2016boundary}. The large-eddy-simulation of the same nozzle with an operation condition $M=0.9$ has been previously validated against the experiments in terms of nozzle-exit conditions. Here the current comparisons serve as a prior validation of these conditions for $M=0.4$ so they are be not analysed further---rather the object of the analysis is the interior nozzle dynamics. 

\begin{table}
	\begin{center}
		\begin{tabular}{cccccccc} 
					
			& $Re$ & $\delta$ & ${\theta}$  & $Re_{\delta}$  & $Re_{\theta}$  & \\[10pt]
			
			Experiment & $4.2 \cdot 10^{5}$ & $9.0 \cdot 10^{-2}$ & $9.5 \cdot 10^{-3}$ & $3.8 \cdot 10^{4}$  &    $4.0 \cdot 10^{3}$  \\[10pt]
			LES & $4.5 \cdot 10^{5}$ & $8.9 \cdot 10^{-2}$   & $8.4 \cdot 10^{-3}$ & $4.0 \cdot 10^{4}$  &    $3.8 \cdot 10^{3}$     \\[10pt]
			
		\end{tabular}
		
		\caption{Nozzle-exit boundary-layer integral parameters for the $M=0.4$ turbulent jet. Experimental results are reported in \citet{cavalieri2013wavepackets}. Boundary-layer thickness $\delta_{99}$ is estimated as the wall-normal distance at which velocity reaches \%99 of local mean centerline velocity $U_c$. Reynolds number based on the integral parameters are defined as $Re_{\delta}={U_{c}{\delta}}/{\nu}$ and $Re_{\theta}={U_{c}{\delta}}/{\nu}$. }\label{t:bl_integral}
		
	\end{center}
\end{table}

The straight circular cross-section and thin boundary layer ($\delta/r<<1)$ allows curvature effects to be neglected and the mean turbulent nozzle boundary layer to be approximated as a flow over a flat plate \citep[see][p. 446]{schlichting2016boundary}. Velocity profiles can thus be compared with results obtained from experimental and numerical data available in the literature. \citet{schlatter2010assessment} computed a canonical flat-turbulent-boundary layer using direct numerical-simulation, providing well-resolved boundary-layer profiles at varying momentum-thickness-based Reynolds numbers, that facilitate evaluation of the current nozzle profiles in terms of inner and outer units. The velocity scale for inner units is given by friction velocity $u_{\tau}/{U_c}=\sqrt{c_f/2}$, where $c_f$ refers to skin-friction coefficient with local nozzle centerline velocity $U_{c}$. The associated length scale simply becomes $\nu/u_{\tau}$ and the distance from the nozzle surface ($y=0.5-r$) and mean velocity scaled in inner units are written respectively as $y^{+}={u_{\tau}} {y}/{\nu}$ and $U^{+}=U/{u_{\tau}}$.



\begin{figure}
	\centering
	\includegraphics[trim=0 5mm 0 0, clip, scale=1]{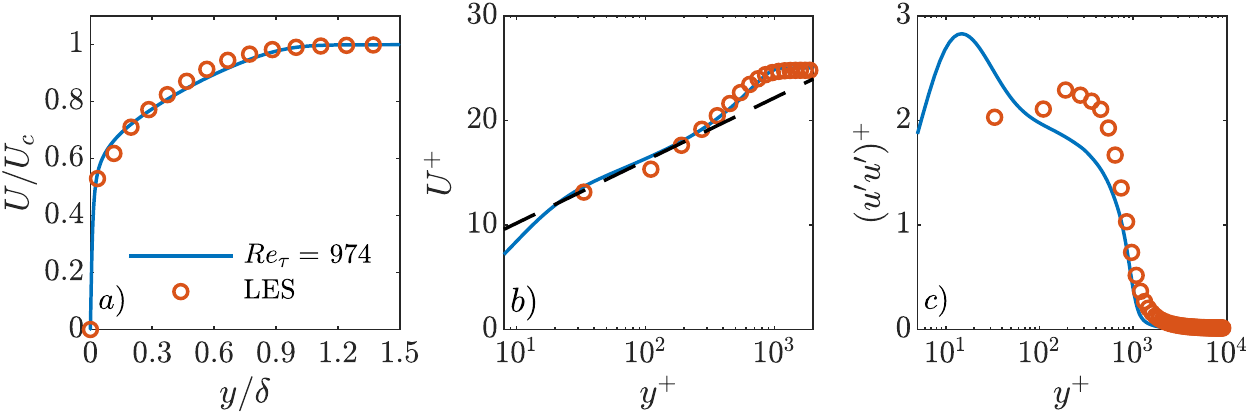}
	\includegraphics[trim=0 0mm 0 0, clip, scale=1]{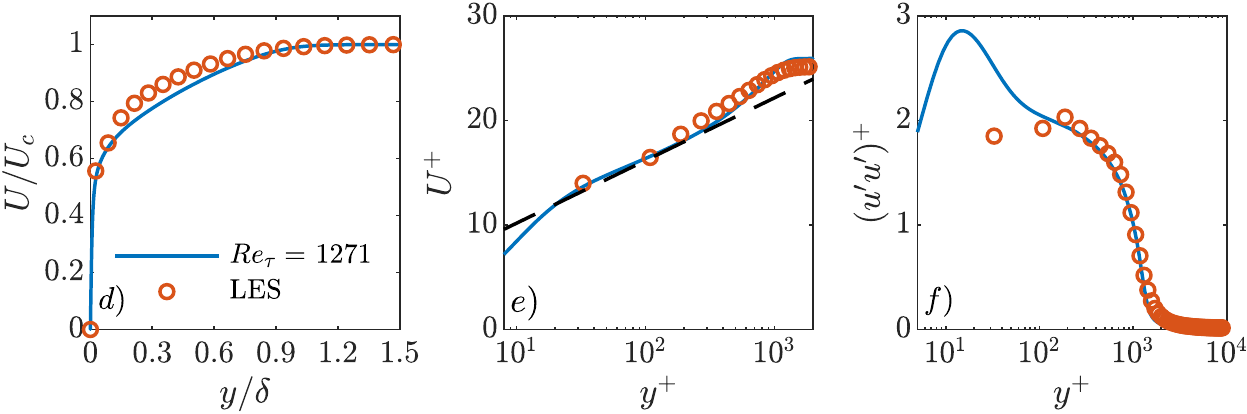}
	\caption{Nozzle boundary-layer profiles at $x=-1.5$ (top row) and  $x=-0.95$ (bottom row): mean velocity in (a,d) outer and (b,e) inner variables; (c,f) streamwise Reynolds stress in inner variables. DNS data is taken from \citet{schlatter2010assessment} at $Re_{\tau}=974$ and $1271$. Dashed lines in (b,e) is logarithmic law $U^{+}={\frac{1}{\kappa}}log(y^{+})+C$ with $\kappa=0.384$ and $C=4.17$.}\label{fig:BLprofiles}
\end{figure}

The mean LES and DNS profiles are compared for two streamwise positions, at which the friction Reynolds numbers $Re_{\tau}={u_{\tau}{\delta}/{\nu}}$ are equivalent. At $x=-1.5$, the LES mean velocity profile already shows reasonable agreement with DNS profile in figure \ref{fig:BLprofiles} and collapses with a slope proportional to $1/\kappa$ in the overlap region---though the variance of streamwise velocity fluctuations is overestimated (figure \ref{fig:BLprofiles}(c)). Further downstream at $x=-0.95$, figure \ref{fig:BLprofiles}(f) shows that LES variance profile compares very well with those of DNS, and likewise mean velocity profile fits reasonably to the power and log laws (figure \ref{fig:BLprofiles}(d,e)). Overall, the comparisons corroborate the idea that the LES nozzle boundary layer develops into a turbulent regime far upstream of the nozzle-exit and its large-scale dynamics are captured by the LES.

\subsection{Azimuthal structure of nozzle flow}	

\begin{figure}
	\centering
	\includegraphics[scale=1]{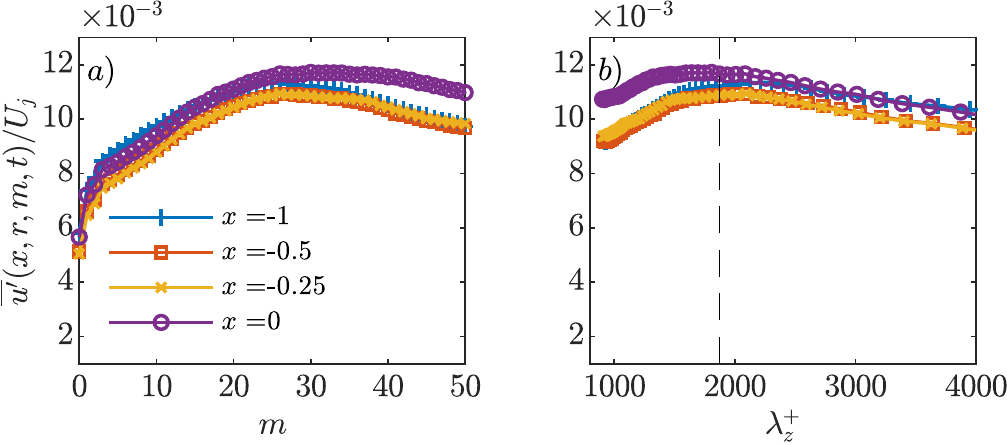}
	\caption{Root-mean square of streamwise velocity fluctuations as a function of azimuthal wavenumber at $r=0.4895$, $y^{+} \approx 185-188$: (a) scaled with jet-radius, (b) in pseudo-spanwise coordinates. Dashed-line indicates 1.5$\delta_{99}$ at $x=-1$.}
	\label{fig:ux_fm}
\end{figure}

The azimuthally periodic fluctuation field $q(x,r,\theta,t)$---containing streamwise, radial and azimuthal velocities  $u_x$, $u_r$, $u_{\theta}$, respectively, density $\rho$ and temperature $T$--- is first decomposed into Fourier modes

\begin{equation}
\label{e:function_azimuthal_decomp}
q(x,r,\theta,t)={\sum_m}{\tilde{q}}_{m}(x,r,t) e^{im\theta}, 
\end{equation}

\noindent where ${\tilde{q}}_{m}$ is the modal amplitude associated with azimuthal wavenumber $m$. Figure \ref{fig:ux_fm} shows the energy distribution of streamwise velocity fluctuations as a function of azimuthal wavenumber along $y^{+} \approx 185$ at several axial positions. It is seen that a large proportion of energy resides predominantly in higher-order azimuthal modes, and the peak is located at  $m\approx 23$ for $x=-1$. Once the spectrum is presented as a function of pseudo-spanwise wavenumber, $\lambda_z=2{\pi}r/m$, the peak scales with boundary-layer thickness $\lambda_z / \delta \approx 1.5$. This matches results for the outer peak at $\lambda_z / \delta \approx 1$ in turbulent boundary layers \citep{eitel2014simulation}. These fluctuations, with azimuthal wavelength of the order of the boundary layer thickness, correspond to turbulent motions referred to as superstructures~\citep{smits2011high}, which are elongated in the streamwise direction. These are the dominant structures in the velocity fluctuations far from the wall. However, with respect to the question we address, low-azimuthal wavenumbers have non-zero amplitudes throughout the nozzle. Since the jet wavepackets related to the peak sound radiation have low azimuthal wavenumber, it is indeed these components of the nozzle flow that we explore, in terms of their connection with downstream wavepackets and the modelling framework that would describe their behaviour within the nozzle.

\subsection{Wavepackets in the nozzle}	

\begin{figure}
	\centering
	\includegraphics[scale=1]{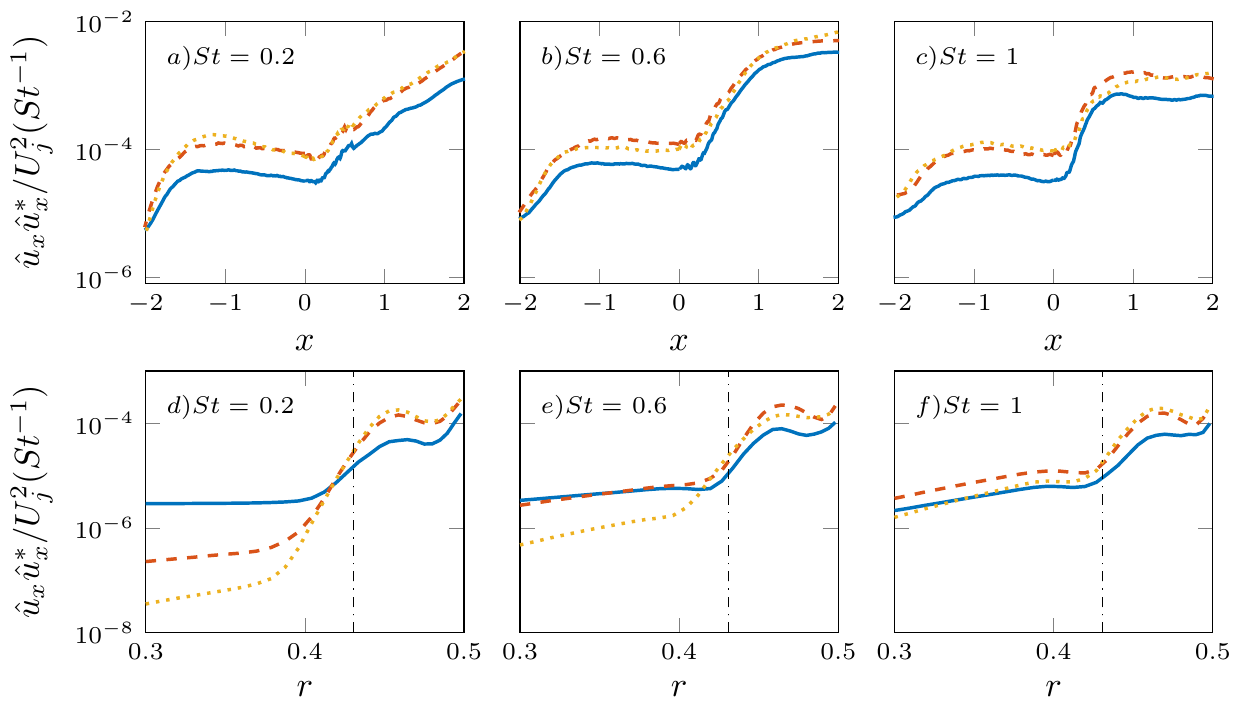}
	\tikzexternaldisable	
	\caption{P.s.d. of streamwise velocity fluctuations for $M=0.4$ jet. (a,b,c) Axial envelopes at $r=0.45$. (d,e,f) Radial profiles at $x=-1$; local boundary thickness (\protect \dashdotted). $m=0$ (\protect\solidline); $m=1$ (\protect\dashedline); $m=2$ (\protect\dottedline).}
	\tikzexternalenable
	\label{fig:PSD_Ur_m02_x010a}
\end{figure}

Under the assumption of ergodicity, different temporal segments of the signal can be treated as statistically independent realisations. The entire time signal is partitioned into 50\% overlapping blocks containing 256 snapshots, providing $N_b=$ 78 independent realisations. Short-time Fourier transforms are applied to each realisation to obtain the Fourier modes ${\hat{q}}_{m\omega}$,

\begin{equation}
\label{e:fourier_time}
{\tilde{q}}_{m}(x,r,t)=\int_{\omega}{\hat{q}}_{m\omega}(x,r) e^{-i\omega t}d\omega,
\end{equation}

\noindent where $\omega=2 \upi St$ refers to angular frequency. A Hann window is applied to time segments to mitigate spectral leakage. Estimates of power-spectral-density (p.s.d.) for a given frequency - azimuthal wavenumber (hereafter $m-\omega$) pair  can be found using

\begin{equation}
\label{e:function_PSD}
{\hat{P}}_{qq}(x,r)=\frac{1}{N_{b}}\sum_{i=1}^{N_{b}} {\hat{q}}_{m\omega}^{(i)}(x,r){\hat{q}}_{m\omega}^{*(i)}(x,r),
\end{equation}

\noindent where ${\hat{q}}_{m\omega}^{(i)}$ is the $i$-th segment, ${\hat{q}}_{m\omega}^{*(i)}$ its complex conjugate, and $N_b$ is the total number of realisations. Figure \ref{fig:PSD_Ur_m02_x010a}(top row) tracks the axial envelope of the streamwise velocity fluctuations of the first three azimuthal modes along $r=0.45$, for three representative frequencies. All $m-\omega$ pairs of interest undergo a similar evolution: fluctuations experience a steep exponential growth in the initial region of the nozzle that ceases approximately around $x=-1$;  there is then a slower amplitude decline---with frequency dependent decay rate---towards the jet-exit plane. Their amplitudes then amplify dramatically downstream of the jet exit. Finally figure \ref{fig:PSD_Ur_m02_x010a}(bottom row) presents the radial structure of the streamwise velocity component of the modes in the nozzle, revealing structures as large as the local boundary-layer thickness. In what follows, we explore whether the dynamics of these low-order nozzle structures are correlated with the downstream jet dynamics.

The foregoing results provide an initial characterisation of the flow dynamics upstream of the nozzle exit, that shows the turbulent nozzle flow to contain low-order azimuthal structures comparable in size to the thickness of the nozzle-boundary-layer. The first three azimuthal modes ($m=0,1,2$), whose downstream jet counterparts are important for sound generation, comprise a small percentage of the total turbulent kinetic energy, similar to what is observed in the jet. We now look to establish whether the dynamics of these low-order nozzle structures are correlated with their counterparts in the free-jet by means of cross-spectral-density. 

Using the Fourier modes obtained from equation \eqref{e:fourier_time}, cross-spectral-density (c.s.d.) between a correlation point ($x^{\prime},r^{\prime}$) and all other points ($x$,$r$) in the flow domain for a given $m-\omega$ pair can be computed as

\begin{equation}
\label{e:function_CSD}
{\hat{C}}_{qq}(x,r,x^{\prime},r^{\prime},m,\omega)=\frac{1}{N_{b}}\sum_{i=1}^{N_{b}} {\hat{q}}_{m\omega}^{(i)}(x,r){\hat{q}}_{m\omega}^{*(i)}(x^{\prime},r^{\prime}).	 
\end{equation}

\begin{figure}
	\begin{subfigure}{\textwidth}	
		\centering
		\hspace*{-0.1cm} 
		\animategraphics[loop,autoplay,scale=1]{12}{plots/Figure5a}{1}{12}
		\vspace{-0.45cm}
		\includegraphics[scale=1]{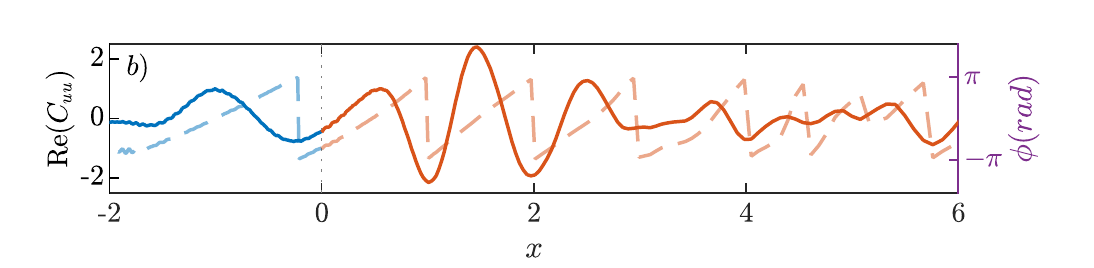}
	\end{subfigure}
	\caption{Cross-spectral-density of axisymmetric streamwise velocity fluctuations with point indicated by '+' for $St=0.6$: (a) c.s.d animation ( ${\color{blue}\blacksquare}\!{\color{Black}\blacksquare}\!{\color{red}\blacksquare}$, $\pm$ 1 of the power-spectral-density at the reference point.); (b) axial envelope on $r=0.45$: Real part (solid line); Phase angle (dashed line). Vertical dotted-line indicates the nozzle-exit. Colour map is courtesy of \citet{kovesi2015good}.}
	\label{fig:CSD_u_m00_St06}
\end{figure}

Figure \ref{fig:CSD_u_m00_St06} shows the c.s.d. of the axisymmetric streamwise velocity field with respect to a reference correlation point within the nozzle boundary-layer for $St=0.6$, highlighting wavepacket activity both upstream and downstream of the jet exit, and significant correlations between these. This, and the axial envelope plotted in figure \ref{fig:CSD_u_m00_St06}b, illustrate a synchronization between structures in the nozzle boundary-layer and jet wavepackets, suggesting that internal nozzle dynamics are casually related to external jet dynamics.

\subsection{Spectral Proper Orthogonal Decomposition (SPOD)}

We now use SPOD as a basis to find coherent structures of low-azimuthal order in the nozzle boundary layer and the downstream region of the turbulent jet, and to further probe the connection between the nozzle and the jet regions. Similar to the previous section, this analysis is concerned with the first three azimuthal modes of the flow field. SPOD provides an orthogonal basis of modes according to their energy at a given frequency. The basis provides an optimal compression in terms of the fluctuation energy with respect to a specified norm. Our SPOD analysis draws on the procedure proposed by \citet{towne2018spectral}, which we briefly outline here.

To perform SPOD, the Fourier realisations obtained in equation \eqref{e:fourier_time} are first stored in a matrix $Q_{m\omega}$ for each \fft pair 

\begin{equation}
\label{e:Fourier} 
{\boldsymbol{Q}}_{m\omega}=\left[{\hat{q}}_{m\omega}^{(1)},  \: {\hat{q}}_{m\omega}^{(2)}, \:  {\hat{q}}_{m\omega}^{(3)}, \: ..... \: \: , {\hat{q}}_{m\omega}^{(N_{b})}\right]. 	 
\end{equation}

\noindent From this, a global two-point cross-spectral-density matrix ${\boldsymbol{S}}_{m\omega}$ is calculated as

\begin{equation}
\label{e:CSD}
{\boldsymbol{S}}_{m\omega}= {\boldsymbol{Q}}_{m\omega} {\boldsymbol{Q}}_{m\omega}\hermconj,
\end{equation}

\noindent with $\left(\cdot\right)\hermconj$ denoting Hermitian transpose. Finally, SPOD modes are obtained through eigendecomposition of the weighted cross-spectral-density matrix, 

\begin{equation}
\label{e:SPOD}
{\boldsymbol{S}}_{m\omega}{\boldsymbol{W}}= {\boldsymbol{\Psi}}_{m\omega} {\boldsymbol{\Lambda}}_{m\omega} {\boldsymbol{\Psi}}_{m\omega}^{-1}.
\end{equation}

\noindent Here, ${\boldsymbol{\Psi}}_{m\omega}$ is the set of SPOD modes for a given $m-\omega$ pair, and its $i^{th}$ column corresponds to the $i^{th}$ SPOD mode ${\psi}^i$. The modes are ranked according to their energies contained in the diagonal eigenvalue matrix ${\boldsymbol{\Lambda}}_{m\omega}=diag(\lambda^1,\lambda^2,...,\lambda^{N_b})$. The mode with the highest energy is referred to as the leading or optimal mode and the lower energy modes are accordingly called sub-optimal modes. ${\boldsymbol{W}}$ is the weight matrix associated with numerical spatial integration and the compressible energy norm \citep{chu1965energy,hanifi1996transient} based on which a inner product is defined. SPOD modes satisfy the orthogonality relation under this weight,

\begin{equation}
{\langle \psi^i,\psi^j \rangle}_{\boldsymbol{W}}={\delta}_{ij}.
\end{equation}

\noindent with $\langle\cdot,\cdot\rangle$ the inner product. As shown by \citet{towne2018spectral}, the modified cross-spectral matrix ${\boldsymbol{\hat{S}}}_{m\omega}= {\boldsymbol{Q}}_{m\omega}\hermconj {\boldsymbol{W}} {\boldsymbol{Q}}_{m\omega}$, has the same non-zero eigenvalues as \eqref{e:SPOD} 

\begin{equation}
\label{e:SPOD_snapshot}
{\boldsymbol{\hat{S}}}_{m\omega}{\boldsymbol{\xi}}_{m\omega}= {\boldsymbol{\xi}}_{m\omega} {\boldsymbol{\tilde{\Lambda}}}_{m\omega}, 
\end{equation}

\noindent and the eigenvectors associated with the non-zero eigenvalues can be obtained through

\begin{equation}
\label{e:SPOD_snapshot_2}
{\boldsymbol{\tilde{\Psi}}}_{m\omega}= {\boldsymbol{Q}}_{m\omega} {\boldsymbol{\xi}}_{m\omega} {\boldsymbol{\tilde{\Lambda}}}_{m\omega}, 
\end{equation}

\noindent which renders the extraction of SPOD modes from large data more straightforward. 

The weight matrix ${\boldsymbol{W}}$ allows extracting SPOD modes to optimally describe dynamics in a specified flow domain \citep{freund2009turbulence}. Two choices are considered in employing this weight. Firstly, we seek modes that maximise the energy downstream of the jet-exit and hence attribute zero weight to the nozzle region. These modes are referred to as `Jet-Weighted' SPOD modes. Subsequently, the region downstream of the jet-exit is excluded to identify energy maximising modes in the nozzle; these are accordingly called `Nozzle-Weighted' SPOD modes.

\subsubsection{Convergence analysis}
In addition to the SPOD analysis with the full data, we also conduct a convergence study by splitting the data into two subsets, each comprising half of the original set, and compute SPOD modes using each subset. A normalised inner product $\beta_{i,j}$ between each SPOD mode $\psi_j$ educed from the full data and $\psi_{i,j}$ from subsets is then computed to evaluate the convergence for every $m-\omega$ pair, 

\begin{equation}
\label{e:function_beta_SPOD}
\boldsymbol{{{{\beta}}_{i,j}}}(m,St)=\frac{\langle \psi_{i,j}, \psi_{j} \rangle}{{\left \| \psi_{i,j} \right \|}_{W} {\left \| \psi_{j} \right \|}_{W}} 
\end{equation}
\noindent where the subscript $i$ denotes the indices of subsets, $j$ corresponds to the $j^{th}$ SPOD mode and ${\left \| \cdot \right \|}$ refers to modulus. This expression leads to correlation coefficients with values between 0 and 1, where $\boldsymbol{\beta}\approx 1$ for a given SPOD mode indicates a statistical convergence, i.e. the SPOD mode maintains its spatial organisation with increasing number of realisations. 

\begin{figure}
	\centering
	\includegraphics[scale=1]{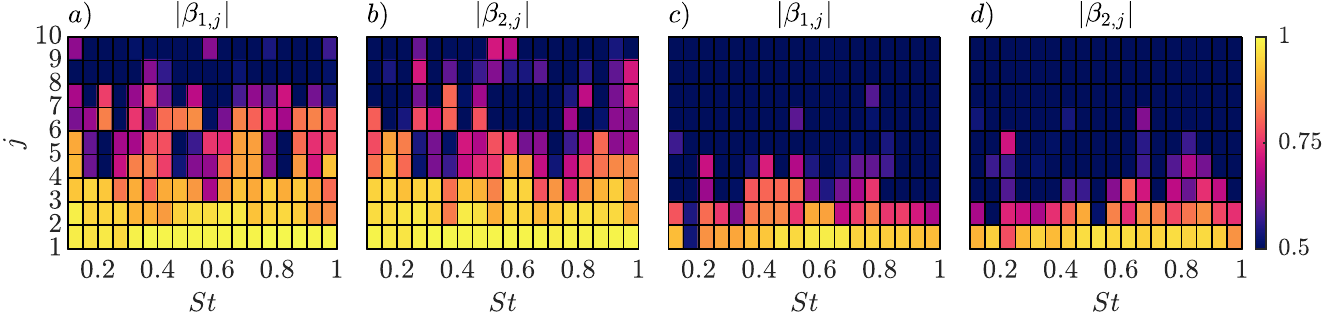}
	\caption{Convergence maps of SPOD modes for $m=0$: (a,b) `Nozzle-Weighted' SPOD modes; (c,d) `Jet-Weighted' SPOD modes.}
	\label{fig:SPOD_convergence_m00}
\end{figure}

Figure~\ref{fig:SPOD_convergence_m00} shows the correlation map obtained for the axisymmetric mode. It shows reasonably converged optimal modes both in `Jet-Weighted' and `Nozzle-Weighted' SPOD analyses, with correlation coefficients greater than 0.8. In the `Nozzle-Weighted' SPOD, the level of agreement is retained for the first two sub-optimal modes for most frequencies; whereas in the `Jet-Weighted' SPOD the correlation is around 0.6-0.7 for first sub-optimal modes, and drops to low levels for higher-order modes. The results altogether indicate that both SPOD analyses provide converged optimal modes for the modes we focus on throughout the paper. We observe similar trends for other $m-\omega$ modes, and provide their convergence maps in Appendix \ref{SPOD_App}.

\subsubsection{Energy spectra}

\begin{figure}
	\centering
	\hspace*{-1cm}
	\includegraphics[scale=1]{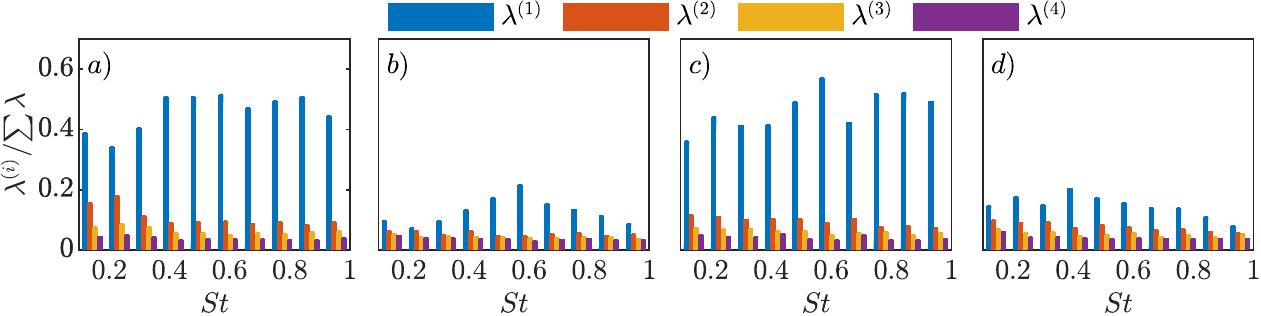}
	\caption{SPOD eigenvalue spectra for `Nozzle-Weighted' (a,c) and `Jet-Weighted' (b,d) as a function of frequency: (a,b) $m=0$; (c,d) $m=1$.}
	\label{fig:SPOD_energies}
\end{figure}

The SPOD eigenvalue spectra resolved as a function of frequency, for the axisymmetric and the first helical mode, is presented in figure \ref{fig:SPOD_energies}. The energies are shown as a fraction of total energy for each \fft. \citet{towne2018spectral} have shown in their SPOD analysis of the same jet, analogous to `Jet-Weighted' SPOD considered here (figure \ref{fig:SPOD_energies}(b,d)), that low-azimuthal dynamics show a low-rank behaviour over a frequency range $0.2<St<2$, i.e. a small number of modes account for the large proportion of the energy at given frequency. In the free-jet region, the associated physical mechanism is mostly due to the inflectional mean-velocity profile that leads to a well-known Kelvin-Helmoltz instability. On the other hand, we observe a low-rank behaviour in a more pronounced way in the `Nozzle-Weighted' SPOD spectrum: the leading SPOD mode has at least more than one-third of the total fluctuation energy over a range of frequencies. It is accordingly anticipated that low-azimuthal nozzle dynamics would be predominantly characterised by this mode alone, and we aim to understand its underlying mechanism. These trends persist also for other azimuthal modes, and their eigenvalue spectra are shown in \S \ref{SPOD_App}.

\subsection{The link between nozzle structures and downstream wavepackets}

\begin{figure}
	\centerline{\includegraphics{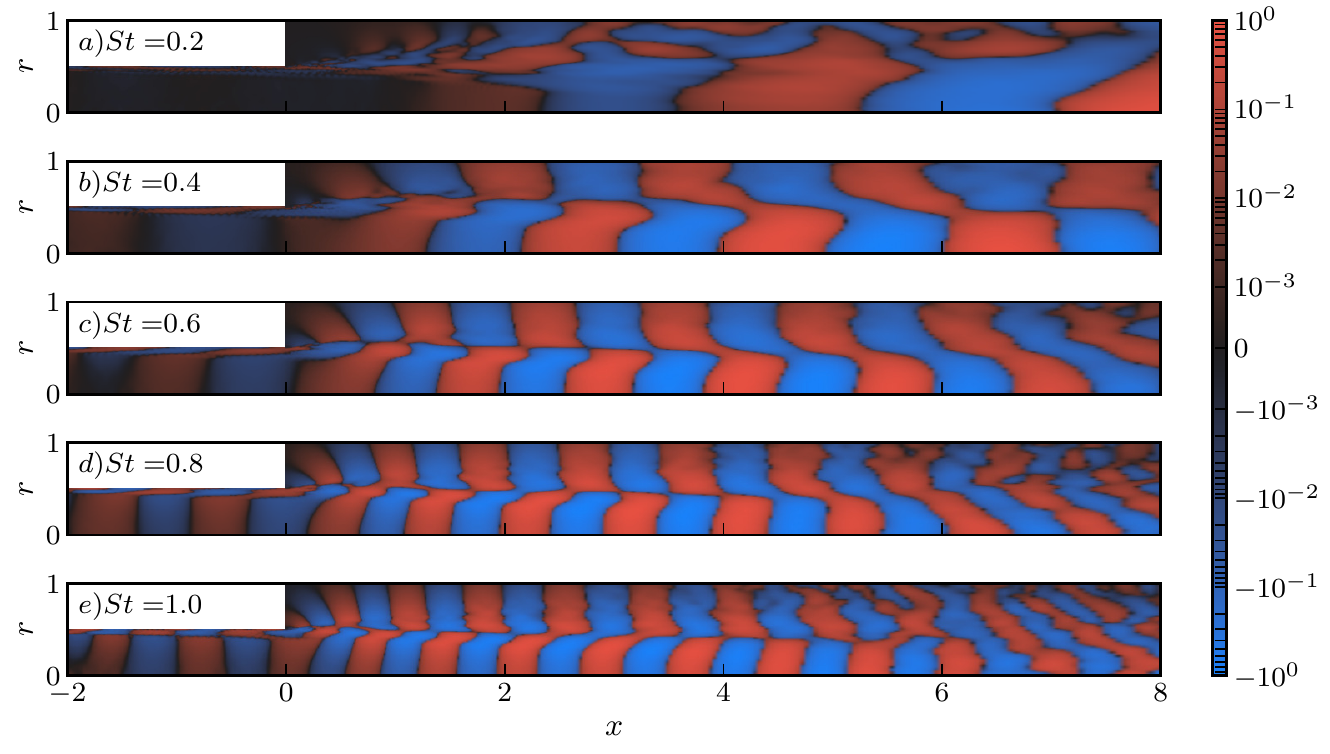}}
	\caption{Streamwise velocity component of the leading `Jet-Weighted' SPOD mode for $m=0$. Each field is normalised with its absolute value in the domain and the contour levels are distributed logarithmically.}
	\label{fig:1stPOD_JW_m00}
\end{figure}

Figure~\ref{fig:1stPOD_JW_m00} shows the streamwise velocity component of the leading axisymmetric `Jet-Weighted' SPOD mode for a range frequencies. Wavepackets are clearly seen in the nozzle boundary layer and further downstream in the jet, and more pronounced compared to CSD, as SPOD filters out uncorrelated fluctuations. Wavepackets in the initial shear-layer are known to be associated with the Kelvin-Helmoltz instability mechanism, and further downstream their dynamics exhibit the characteristics of non-modal evolution following the Orr mechanism \citep{tissot2017wave,tissot2017sensitivity, schmidt2018spectral,lesshafft2019resolvent}. These wavepackets appear to be coupled with wavepackets mainly supported within the nozzle boundary layer, which are clearly seen to tilt into the mean-flow direction for $St=1$ (figure \ref{fig:1stPOD_JW_m00}e). Wavepackets in both regions are characterised by an increasing wavenumber with increasing frequency: this explains the poor wavepacket eduction within the nozzle at low frequencies both in SPOD and CSD (e.g. $St=0.2$), for which wavelengths extend up to approximately five to six jet diameters from the nozzle-exit plane.

\begin{figure}
	\centerline{\includegraphics{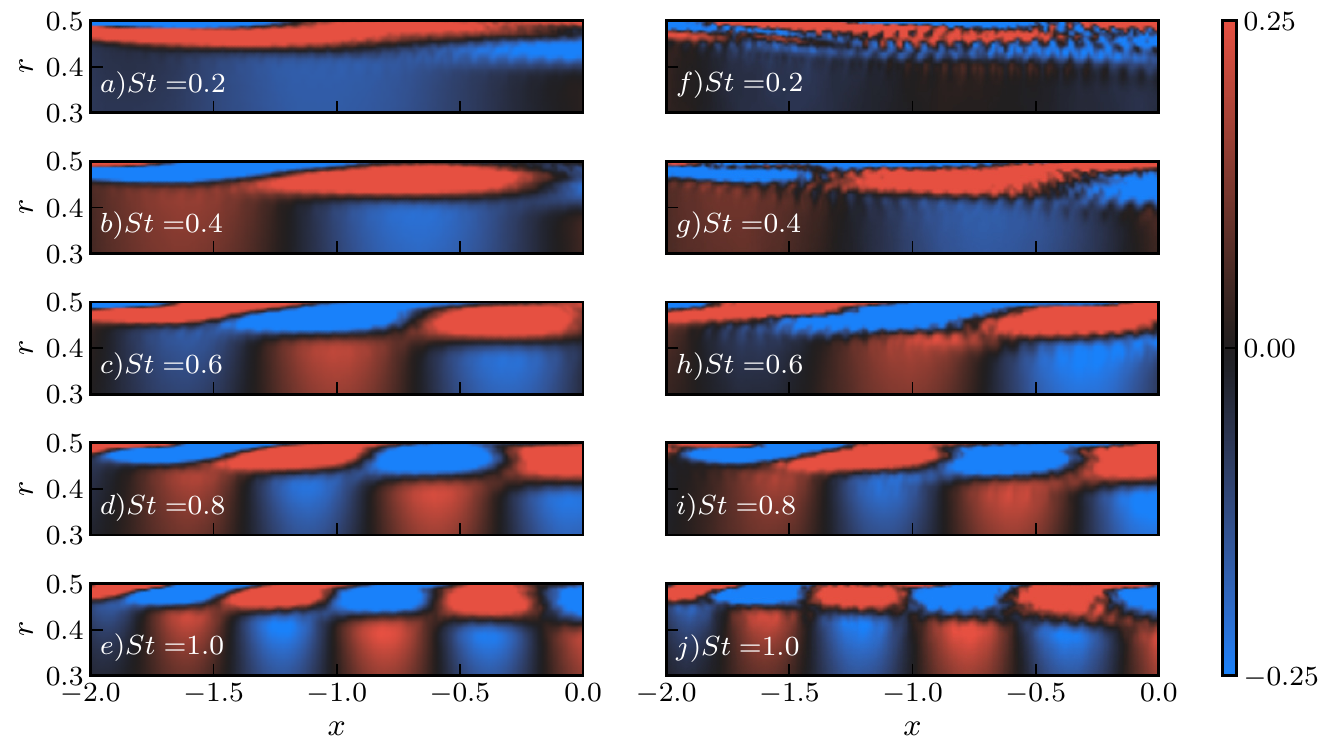}}
	\caption{Streamwise velocity component of the leading SPOD mode in the nozzle at different frequencies for $m=0$: (a-e) `Nozzle-Weighted' SPOD; (f-j) `Jet-Weighted' SPOD. Each field is normalised with its absolute value in the nozzle.}
	\label{fig:POD_JWvsNW_m00}
\end{figure}

\begin{figure}
	\centerline{\includegraphics{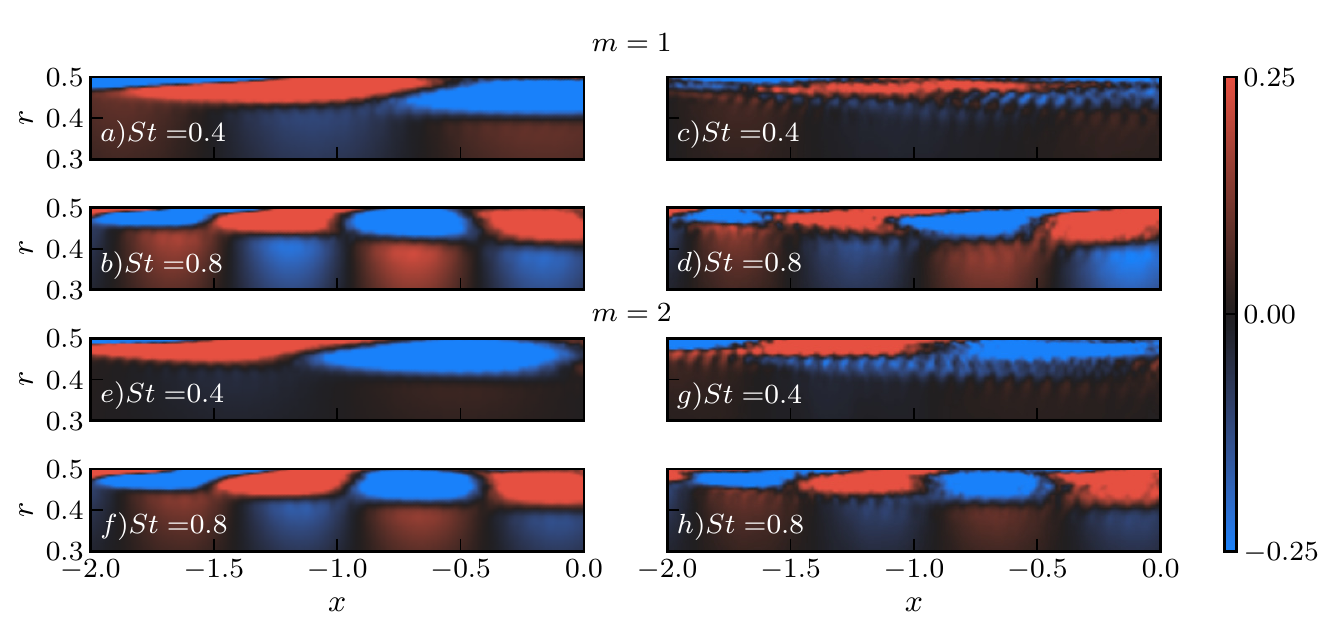}}
	\caption{Streamwise velocity component of the leading SPOD mode in the nozzle at different frequencies for $m=1$ (a-d) and $m=2$ (e-h): `Nozzle-Weighted' SPOD (left column); `Jet-Weighted' SPOD (right column). Each field is normalised with its absolute value in the nozzle.}
	\label{fig:POD_NWvsJW_m01to02}
\end{figure}

To determine whether the nozzle structures associated with jet wavepackets---educed via `Jet-Weighted' SPOD--- correspond to the most-energetic axisymmetric nozzle structures, figure \ref{fig:POD_JWvsNW_m00} compares the streamwise velocity fields of the leading `Jet-Weighted' (taken from figure \ref{fig:1stPOD_JW_m00}) and `Nozzle-Weighted' SPOD modes within the nozzle. This comparison clearly reveals, over a range of frequencies, that the leading axisymmetric SPOD modes within the nozzle, which the `Jet-Weighted' SPOD modes show to be connected to the downstream axisymmetric jet dynamics, are also the most-energetic axisymmetric nozzle structures. Moreover, a similar conclusion can be made where the higher-order azimuthal modes are concerned (figure \ref{fig:POD_NWvsJW_m01to02}). Differences between the fields appear mainly as noise for the `Jet-Weighted' modes inside the nozzle. Such modes have lower amplitudes inside the nozzle, as shown in figure \ref{fig:1stPOD_JW_m00}, and this region is thus more prone to convergence issues. The dominant features inside the nozzle are nonetheless captured in a similar manner by both `Nozzle-Weighted' and `Jet-Weighted' SPOD modes.

The SPOD modes inside the nozzle have a wave behaviour. The phase speed, studied in detail in \S \ref{resolvent_nozzle}, is lower than the jet velocity, which characterises a predominantly hydrodynamic phenomenon. Such waves are similar to the spanwise coherent structures in the turbulent boundary layer over an airfoil, extracted by SPOD by \citet{sano2019trailing}. As in the cited work, the streamwise velocity displays a change of phase in the wall-normal direction. Despite the low overall energy of azimuthal modes $m=0,1,2$ in the nozzle (as seen from figure \ref{fig:PSD_Ur_m02_x010a}), such coherent waves inside the nozzle are clearly connected to the downstream jet wavepackets, as seen in the SPOD modes displayed in figures \ref{fig:1stPOD_JW_m00}-\ref{fig:POD_NWvsJW_m01to02}.

We thus establish a connection between downstream wavepackets and structures observed in the turbulent nozzle boundary layer. It appears that where the nozzle region is concerned, the structures that dominate its low-order dynamics are also those that connect with the downstream wavepackets. Local linear spatial stability equations are used to model the nozzle boundary-layer structures educed using the `Nozzle-Weighted' SPOD in the following section. Here and throughout, the results concerning this SPOD are reported and the prefix `Nozzle-Weighted' is dropped.

\section{Linear stability analysis}\label{lst_nozzle}

We consider the linear spatial stability of incompressible disturbances on a locally parallel mean flow in a circular nozzle. We consider the flow as incompressible on account of the fact that the jet is isothermal and that density variation in the nozzle is almost negligible. Moreover, the compressibility effects on the stability characteristics of turbulent jets play a little role at the $M$ number of the studied jet \citep[fig. 17]{michalke1984survey}  . Homogeneity of the mean flow in time, streamwise and azimuthal directions permits the use of normal modes

\begin{equation}
\tilde{q}(x,r,\theta,t)=\hat{q}(r)e^{i({\alpha} x + m {\theta}-{\omega} t)},
\end{equation}

\noindent where $q$ is the state vector contaning the flow variables--- $u_x$,$u_r$,$u_{\theta}$,and $p$ denote streamwise, radial, azimuthal velocities and pressure; $m$ and $\alpha$ are azimuthal and streamwise wavenumbers, and $\omega$ refers to angular frequency. Following the substitution of the perturbation ansatz into the linearised, incompressible Navier-Stokes equations, the equation system can be recast as a generalised eigenvalue problem as

\begin{equation}\label{eq:lst_eig}
\mathcal{L}\hat{q}_{m\omega}=\alpha\mathcal{F}\hat{q}_{m\omega},
\end{equation}

\noindent whose solution for a given Reynolds number, real angular frequency $\omega$ and azimuthal wavenumber $m$ results in eigenmodes $\hat{q}_{m\omega}(r)$ with associated complex eigenvalues $\alpha$. The operators in discrete form of equation \eqref{eq:lst_eig} are provided in Appendix \ref{Operators_LST}. Spatial growth or decay of a stability eigenmode is governed by the imaginary part of the streamwise wavenumber: if ${\alpha}_i<0$ the disturbance will grow exponentially in the positive streamwise direction, whereas it will experience a decay if ${\alpha}_i>0$. The complete form of the equation \eqref{eq:lst_eig} includes ${\alpha}^2$ viscous terms associated with second-order axial derivatives of perturbations, which are excluded
in the current analysis, similar to what has been done in \citet{rodriguez2015study} and \citet{gudmundsson2011instability}. \citet{li1997spectral} have shown that excluding these terms removes the upstream traveling vorticity and entropy waves from the spectrum, and has negligible effect on other branches. This is also verified in the current work; omitting these terms does not influence the boundary-layer and free-stream modes we consider.

A pseudospectral method is used for discretisation of the system of equations. This results in differentitation matrices for derivatives along the radial direction. Chebyshev collocation points are used to discretise the computational domain, with a numerical grid spanning from the nozzle centerline up to the nozzle-wall ($0.5\geq r >0$). The convergence of the boundary-layer and free-stream eigenmodes is achieved with $N_c=181$ discretisation points. The equations are completed with appropriate boundary conditions. On the wall ($r=0.5)$, no-slip condition is applied for three velocity components

\begin{equation}
{\hat{u}}_{x}={\hat{u}}_{r}={\hat{u}}_{\theta}=0. 
\end{equation}

\noindent At the jet centerline ($r\rightarrow 0$), particular treatment is required to obtain a bounded solution. For an incompressible flow, these conditions are applied following \citet{khorrami1989application} and \citet{lesshafft2007linear}

\begin{center}
	\begin{align}
	&m=0 &: \quad \quad &\hat{u}_{x}^{\prime}=\hat{u}_{r}=\hat{u}_{\theta}=0, &  \hat{p}^{\prime}&=2 \hat{u}_{r}^{\prime \prime} / Re \\
	&{|m|=1} &: \quad \quad &\hat{u}_{x}=\hat{p}=\hat{u}_{r}^{\prime}=0, & \hat{u}_{\theta}&=i \hat{u}_{r} \\
	&{|m|>1} &: \quad \quad &\hat{u}_{x}=\hat{u}_{r}=\hat{u}=\hat{p}=0 . 
	\end{align}
\end{center}

\noindent The governing equations are made non-dimensional using the local mean nozzle centerline velocity ${U^{\ast}_{c}}$ and displacement thickness of the boundary layer $\delta_{1}^{\ast}$ at a given axial position $x_0^{\ast}$: this leads to Reynolds number of $Re_{{\delta}_{1}}={U^{\ast}_{c}}{\delta}_{1}^{*}/{\nu}$. Using these characteristic scales, angular frequency $\omega$, streamwise wavenumber $\alpha$ and velocity simply become

\begin{equation}
U=\frac{U^{*}}{U^{*}_{c}}, \qquad
{\delta_{1}}={\delta^{*}_{1}/D^{*}}, \qquad
\omega=2\upi St {\delta^{x_0}_{1}}, \qquad 
\alpha_{D}={\alpha}^{*}D^{*}, \qquad \alpha=\alpha_{D}{\delta^{x_0}_{1}}.
\end{equation}

\noindent Note that the Strouhal number is based on the nozzle diameter $D^{*}$ and the jet-exit centerline velocity $U_{j}^{*}$, i.e. $St=f^{*}D^{*}/U_{j}^{*}$. The turbulent mean nozzle flow is approximated by a velocity function for zero-pressure-gradient boundary-layer flows \citep{monkewitz2007self}. The composite function can be simply written in terms of wall distance in inner units and displacement-thickness-based Reynolds number,

\begin{equation}
U=f(y^{+},Re_{{\delta}_{1}}).
\end{equation}

\noindent The analytical velocity function ensures well-behaved and smooth radial derivatives and is provided in Appendix \ref{App_VelocityFunction}. Figure~\ref{fig:Meanprofiles} presents comparison between the fit and the LES mean nozzle flow at several streamwise positions.

\begin{figure}
	\centering
	\includegraphics[scale=1]{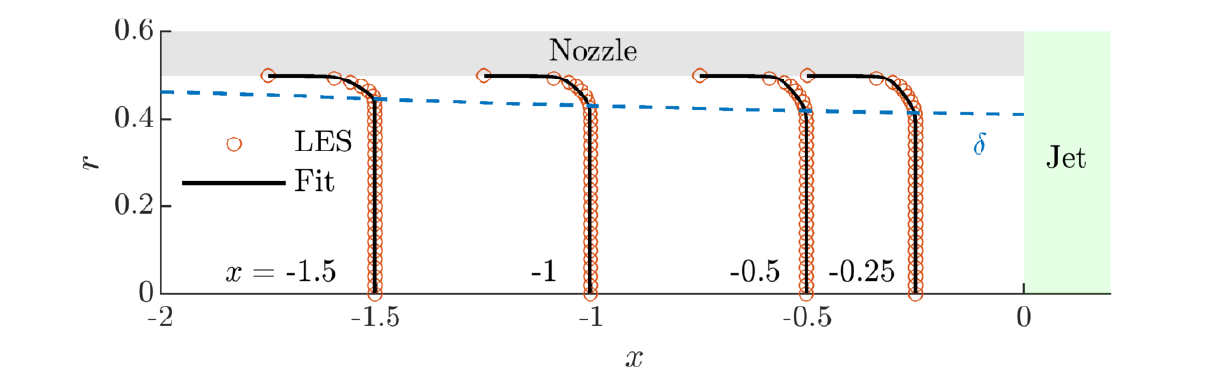}
	\caption{Streamwise mean flow profiles in the nozzle. $U/U_{c}$ is plotted with $U_c$ as local nozzle centerline velocity. Black solid lines are velocity profiles obtained through analytical fit of \citet{monkewitz2007self}.}
	\label{fig:Meanprofiles}
\end{figure}

\subsection{Eigenvalue spectra}	

\begin{figure}
\begin{subfigure}{0.65\textwidth}
	\includegraphics[scale=1]{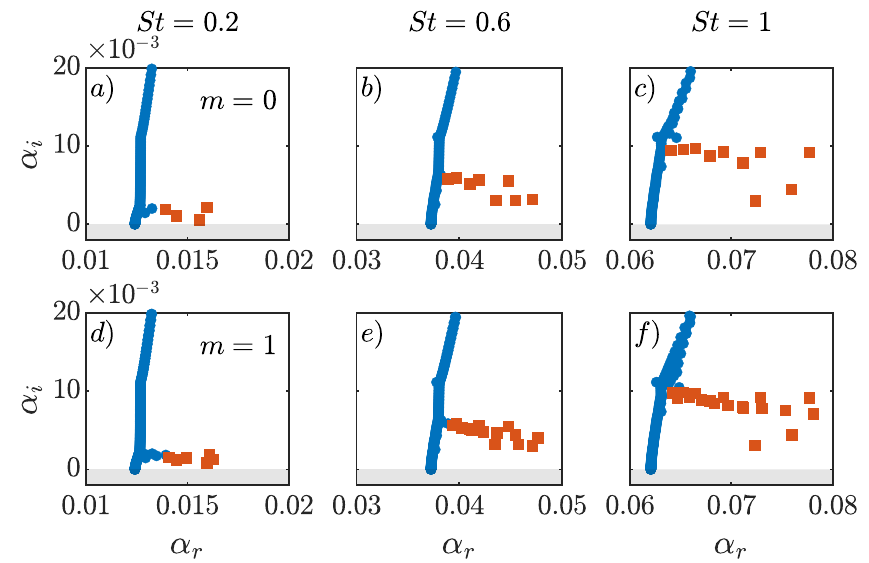}
\end{subfigure}
\begin{subfigure}{0.3\textwidth}
	\includegraphics[width=\textwidth]{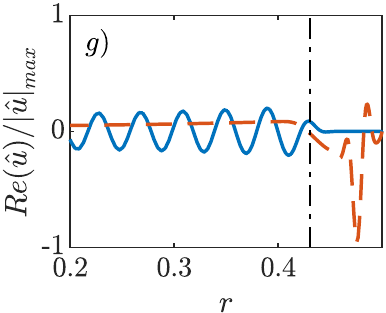}
\end{subfigure}
\tikzexternaldisable
\caption{Eigenvalue spectra at $x_0=x=-1$ for (a-c) $m=0$ and (d-f) $m=1$. \protect\orangesquare Boundary-layer modes; \protect\bluecircle Free-stream modes. g) Real part of streamwise velocity: eigenmodes from free-stream (solid line) and boundary-layer (dashed line) branches.} 
\tikzexternalenable
\label{fig:EigSpec_m00_x10}
\end{figure}

Since the mean boundary-layer profiles at different axial positions are qualitatively similar, solution of the eigenvalue-problem in different turbulent regions of the nozzle produces nearly indistinguishable eigenspectra. The first quadrant of the spectra for axisymmetric disturbances at $x=-1$ is shown in figure \ref{fig:EigSpec_m00_x10}. Eigenvalues can be readily classified into two families: free-stream and boundary-layer modes, and both branches are stable. Modes in the free-stream branch lie roughly on a line parallel to the ${\alpha}_i$ axis, with a phase speed $c_p={\omega/k}$ approximately equal to the mean centerline velocity. These modes are advected by uniform mean flow and eventually damped by viscosity in the streamwise direction \citep{grosch1978continuous}. The boundary-layer branch consists of a cluster of modes with varying phase speeds. The streamwise velocity component of the least-stable eigenmodes from each branch are plotted in figure \ref{fig:EigSpec_m00_x10}g.

Besides the class of eigenmodes introduced above, the axisymmetric eigenmodes can be further regrouped into two families: pairs of \textit{meridional} and \textit{torsional} modes \citep{boberg1988onset,tumin1996receptivity}. A meridional mode has $\hat{u}_{x},\hat{u}_{r},\hat{p}$ components with non-zero amplitudes and azimuthal velocity component $\hat{u}_{\theta}$ equal to zero. On the other hand, a torsional mode has ${\hat{u}}_{\theta}$ with non-zero amplitude, while the other velocity and pressure components are equal to zero. The torsional modes appear due to the decoupling of the azimuthal momentum conservation equation from other equations for $m=0$. However, since the axisymmetric disturbances have almost negligible azimuthal velocity amplitudes in the LES data, these modes are not of major interest in the present study and are excluded from the analysis.

\subsection{Modal decomposition of nozzle wavepackets}	

The solution of equation \eqref{eq:lst_eig} yields non-orthogonal eigenfunctions that constitute a complete basis \citep{salwen1981continuous,tumin1983spatial}, enabling decomposition of the radial profile of a leading SPOD mode $\boldsymbol{\hat{q}_{m\omega}(r)}$ as a linear combination of stability eigenmodes

\begin{equation}
\boldsymbol{\hat{q}_{m\omega}(r)}=\sum_{i}^{N}{{{a}^{(i)}_{{m\omega}}}}{{\hat{q}_{{m\omega}}^{{(i)}}}}(r),
\label{e:pod_r_expansion_lst}
\end{equation}

\noindent with ${{\hat{q}_{{m\omega}}^{{(i)}}}}$ stability eigenmodes and associated complex-valued amplitudes ${{{a}^{(i)}_{{m\omega}}}}$, and $N$ is the order of expansion. Note that the expansion is evaluated for each \fft pair, hence the subscript $\left(\cdot\right)_{m{\omega}}$ is dropped for brevity. 

The eigenfunctions have arbitrary amplitudes, i.e. $a$ is unknown a priori, thus their true amplitudes must be determined. These can be straightforwardly found by using the biorthogonality relation $\langle \hat{\xi}^{(i)},\mathcal{F}\hat{q}^{(j)} \rangle=\delta_{ij}$, where $\hat{\xi}$ denotes eigenfunctions of adjoint system ${\mathcal{L}}^{\dagger}\hat{\xi}^{\dagger}={\alpha}^{\dagger}{\mathcal{F}}^{\dagger}\hat{\xi}^{\dagger}$ and $\delta_{ij}$ the Kronecker symbol.  However, the low resolution very-near the wall in the LES data does not permit the correct calculation of the amplitudes associated with eigenfunctions used in the expression \eqref{e:pod_r_expansion_lst}.  Instead, orthogonal projection can be considered to calculate expansion coefficients \citep{passaggia2009control}. Following the methodology outlined in \citet{alizard2011modeling}, first an orthogonal basis is derived from an LST subspace of dimension $N$, comprising certain non-orthogonal stability eigenmodes, using a Gram-Schmidt method. The radial structure of the leading SPOD mode can then be expanded in terms of orthogonal linear modes as

\begin{equation}
\boldsymbol{\hat{q}(r)}=\sum_{i}^{N}{\Lambda}_{i}{\hat{q}_{{\bot}}^{{(i)}}}(r),
\end{equation}

\noindent where $\hat{q}_{\bot}$ denotes modes of the orthogonal basis with associated expansion coefficients $\Lambda$. The coefficients in the orthogonal basis can be calculated through

\begin{equation}
\label{inner_product}
{\Lambda}_{i}=\langle {\boldsymbol{\hat{q}},{\hat{q}_{{\bot}}^{{(i)}}}} \rangle= \int_{0}^{1}  ({\hat{q}_{{\bot}}^{{(i)}}})^{*}(r) {\boldsymbol{\hat{q}(r)}}rdr,
\end{equation}

with $\langle\cdot,\cdot \rangle$ as Hermitian inner product. The final step of the procedure consists in recovering the true amplitudes of non-orthogonal linear modes of the truncated LST basis, ${\boldsymbol{{a}}}=[a_1,a_2,....,a_n]^T$

\begin{equation}
{\boldsymbol{{a}}}={\mathcal{P}}^{-1} \mathbf{\Lambda},
\end{equation}

\noindent where $\mathbf{\Lambda}=[{\Lambda}_{1},{\Lambda}_{2},....,{\Lambda}_{N}]^T$ and ${\mathcal{P}}_{ij}=<{\hat{q}}^{(i)},{\hat{q}_{{\bot}}^{(j)}}>$. One of the central aims of the current analysis is to find the appropriate order of expansion to accurately describe the leading SPOD mode. Different numbers of eigenmodes are considered to restrict the complete eigenbasis and to make the projection for each $m-{\omega}$ pair: (1) Single mode representation; (2) All of the boundary-layer modes; (3) Boundary-layer and free-stream modes. Decompositions of the LES data with truncated eigenbases are referred to as \textit{rank-1}, \textit{rank-B} and \textit{rank-BF} models respectively.

\begin{figure}
	\centering
	\includegraphics[scale=1]{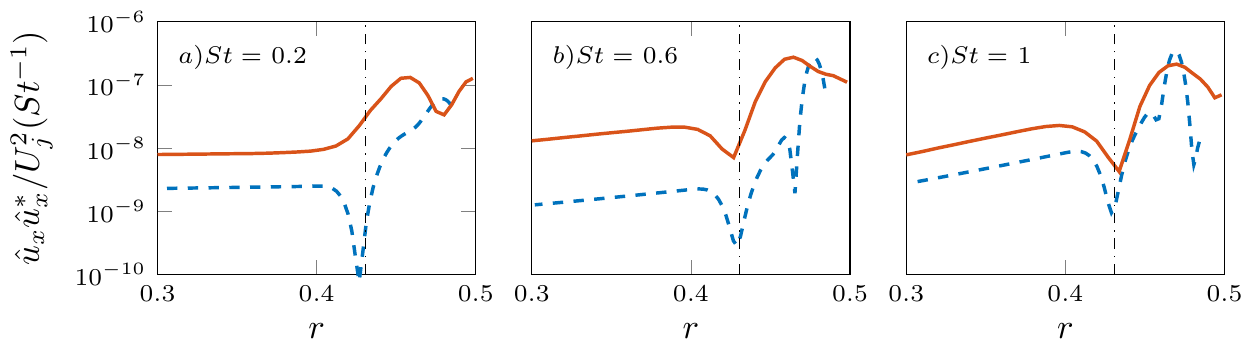}
	\caption{Radial profiles of the streamwise velocity ($m=0$) at $x=-1$: The leading SPOD mode (red solid line) and rank-1 stability model (blue dashed line).}
	\label{fig:U_EigProj_m0_r_x15_ls}
\end{figure}

The leading axisymmetric SPOD mode is first decomposed in terms of the least-stable boundary-layer mode, and the reconstruction is shown in figure \ref{fig:U_EigProj_m0_r_x15_ls}. This shows that a \textit{rank-1} model leads to a poor description of the axisymmetric nozzle structures for all frequencies considered. A global velocity field can be reconstructed using the stability eigenmodes in frequency space as

\begin{figure}
	\centering
	\includegraphics[scale=1]{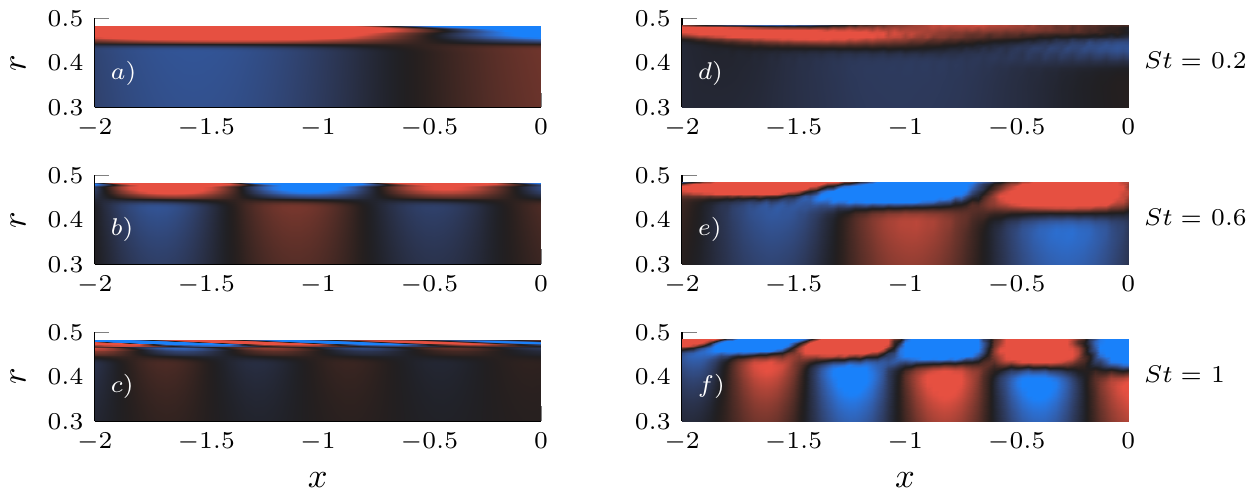}
	\caption{Real component of the streamwise velocity $(m=0)$ within the nozzle. Left: Rank-1 stability model; Right: Leading SPOD mode. Initial condition of the modal reconstruction is determined from orthogonal projection of the leading SPOD mode onto least-stable eigenmode at axial station $x=-1.5$ (${\color{blue}\blacksquare}\!{\color{Black}\blacksquare}\!{\color{red}\blacksquare}$, $\pm$  0.25 of the maximum absolute value in the nozzle).}
	\label{fig:Global_U_EigProj_m00_ls}
\end{figure}

\begin{equation}
\boldsymbol{{\tilde{q}}_{m\omega}}(x,r)=\sum_{i}^{N}{a_{i} {{\hat{q}}^{(i)}_{m\omega}(r)} e^{j\alpha^{(i)} (x-x_0)}},
\label{eq:spatial_recons}
\end{equation}

\noindent where $\alpha^{(i)}$ is the eigenvalue of a given stability eigenmode obtained at axial location $x_0$. This expression allows each mode, with its amplitude computed at a given axial location, radial shape and phase, to contribute to the spatial reconstruction that will be compared to the leading SPOD mode within the nozzle. Figure \ref{fig:Global_U_EigProj_m00_ls} shows, consistent with the poor agreement in radial shape, that the \textit{rank-1} model differs from the data in spatial organisation: the disagreement is particulary pronounced at $St=1$, whereas for other frequencies, while the model captures some of the observed features, reconstructed wavepackets display pronounced differences, such as a lower wavelength than what is observed in the SPOD mode. \textit{Rank-1} models using other modes of the boundary-layer family result in similarly poor agreement, indicating that a higher rank model is necessary. 

\begin{figure}
	\centering
	\includegraphics[scale=1]{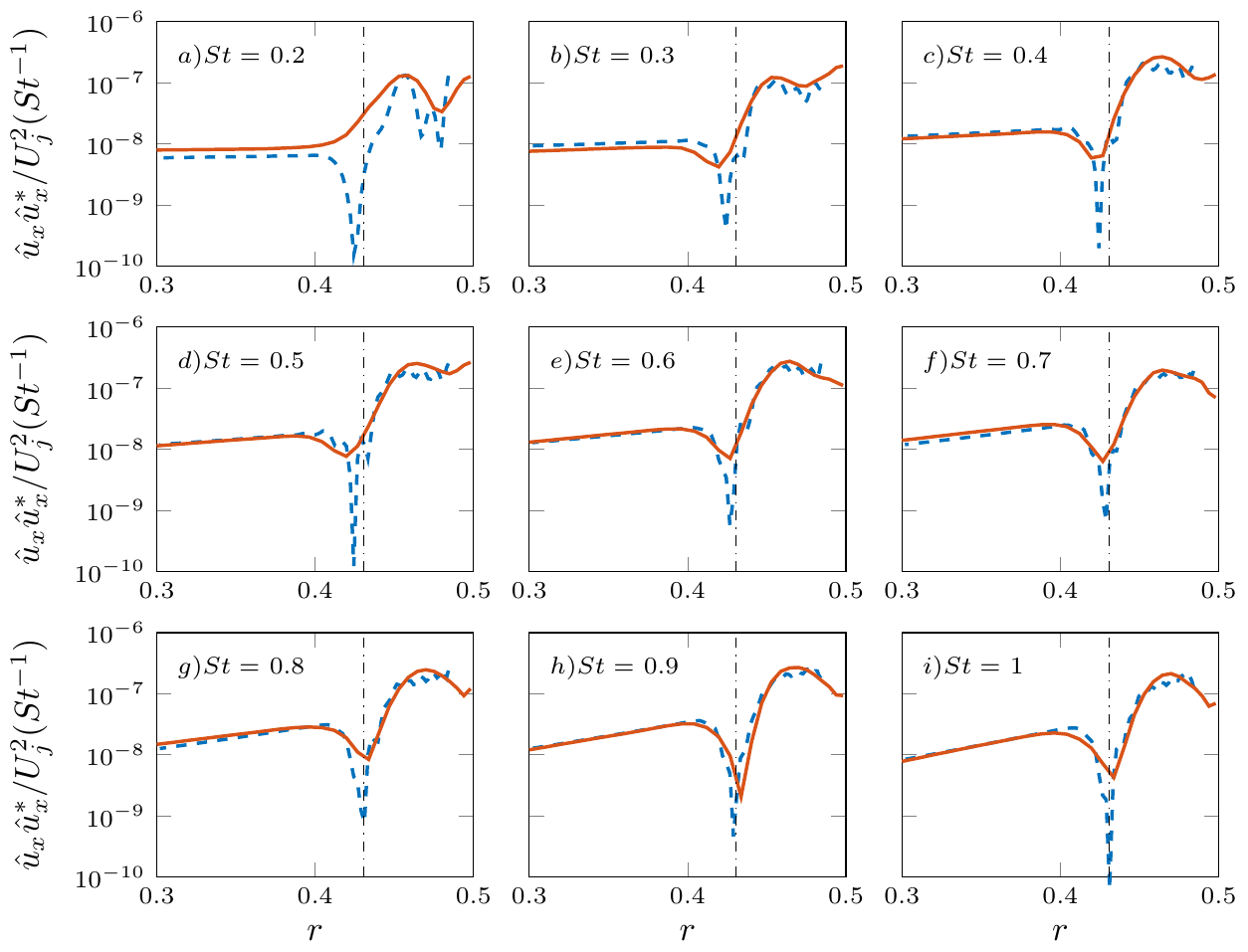}
	\caption{Radial profiles of the streamwise velocity ($m=0$) at $x=-1$: The leading SPOD mode (red solid line) and \textit{rank-B} stability model (blue dashed line). Dotted line is the local boundary layer thickness. Note that the order of the stability model $N$ is frequency dependent: $N_{min}=3$ and $N_{max}=10$ for $St=0.2$ and $St=1$ respectively.}
	\label{fig:U_EigProjvsPOD_m00_wp_x-10_rankBLB}
\end{figure}

\begin{figure}
	\centering
	\includegraphics[scale=1]{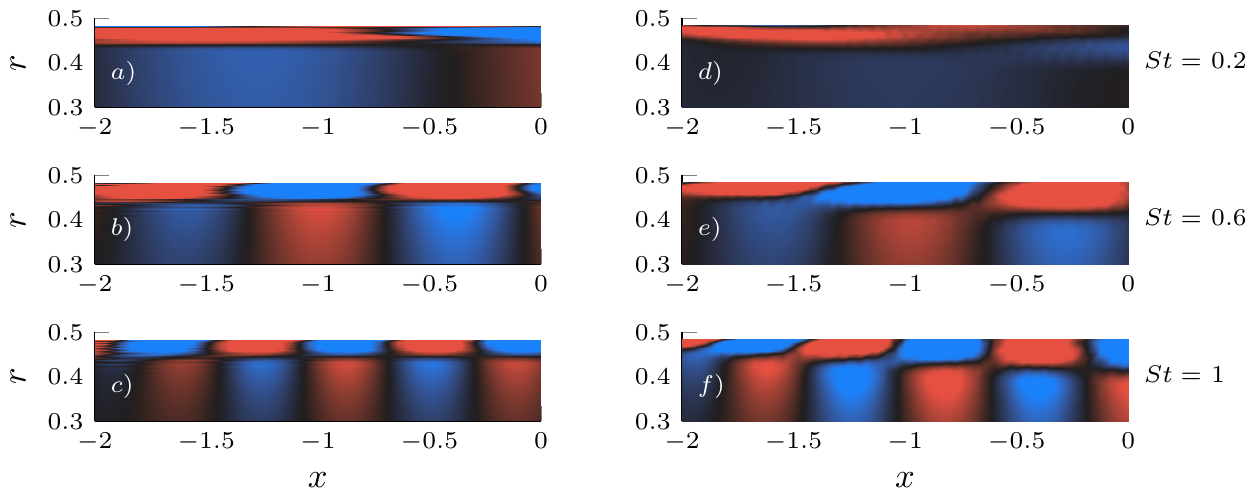}
	\caption{Spatial fields of axisymmetric velocity fluctuations $(m=0)$ within the nozzle: (a,b,c) \textit{rank-B} stability model; (d,e,f) leading SPOD mode. Real part is shown.}
	\label{fig:Global_U_EigProj_m00}
\end{figure}

\begin{figure}
	\centering
	\includegraphics[scale=1]{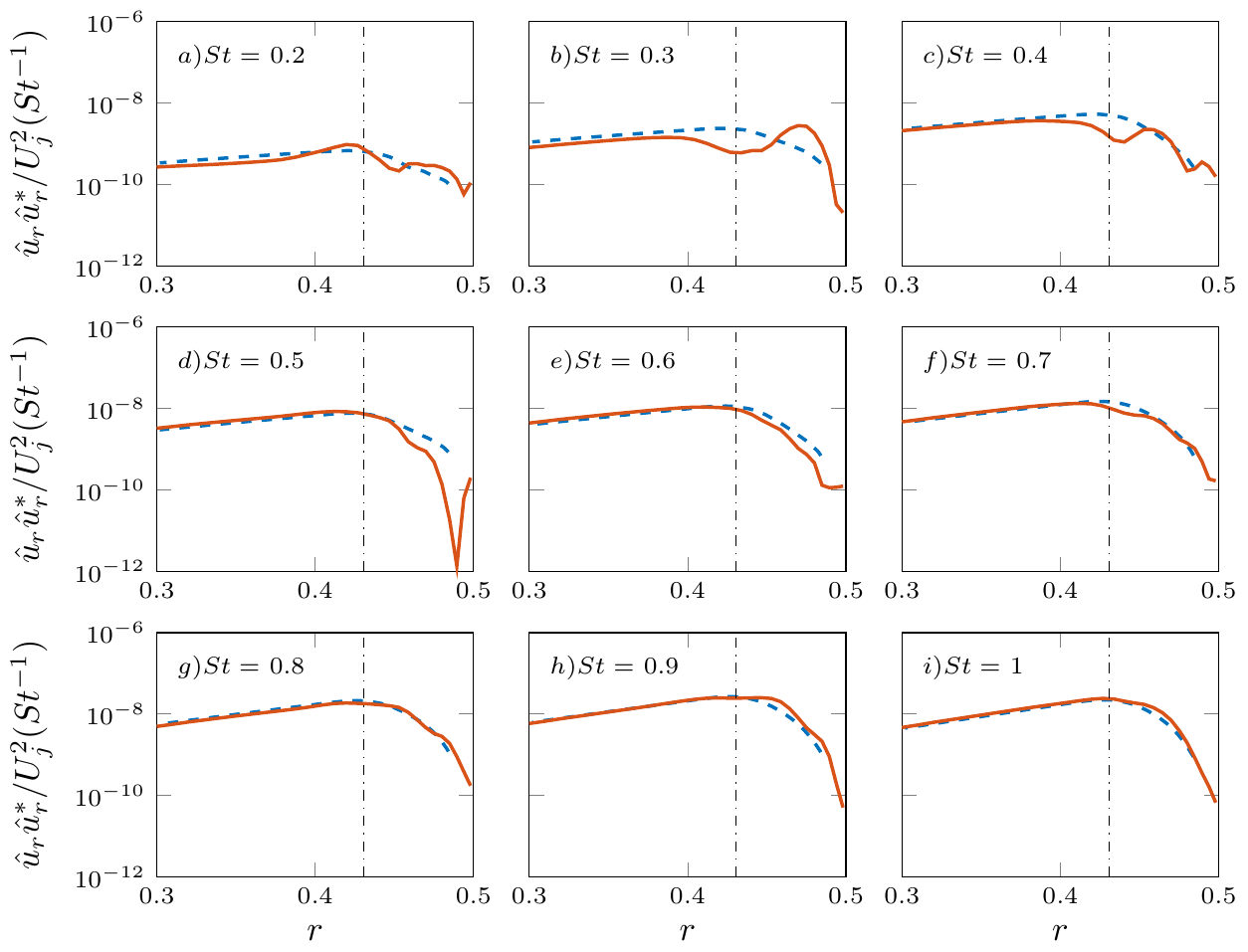}
	\caption{Comparison of radial velocity ($m=0$) at $x=-1$: The leading SPOD mode (red solid line) and \textit{rank-B} stability model (blue dashed line). Dotted line is the local boundary layer thickness.}
	\label{fig:V_EigProjvsPOD_m00_wp_x-10_rankBLB}
\end{figure}

Using the entire boundary-layer branch in the expansion provides a second low-rank description (\textit{rank-B}). The number of eigenmodes $N$ used in the \textit{rank-B} expansion depends on the frequency considered: e.g the order is $N=4$ for $St=0.1$ and $N=10$ for $St=1$ for axisymmetric structures. Figure \ref{fig:U_EigProjvsPOD_m00_wp_x-10_rankBLB} shows a comparison of streamwise velocity fluctuation profiles between the nozzle structures and the \textit{rank-B} wavepacket model. Agreement has considerably improved, the radial form of the nozzle boundary-layer structures is recovered with the low-rank stability model. Figure \ref{fig:V_EigProjvsPOD_m00_wp_x-10_rankBLB} shows the radial velocity profiles at the same axial position, and the stability model compares successfully with the nozzle wavepackets educed via SPOD over a range of frequencies. The comparisons in spatial organisation (figure \ref{fig:Global_U_EigProj_m00}) also show substantial agreement between the \textit{rank-B} stability reconstruction and the leading SPOD mode.

\begin{figure}
	\centering
	\includegraphics[scale=1]{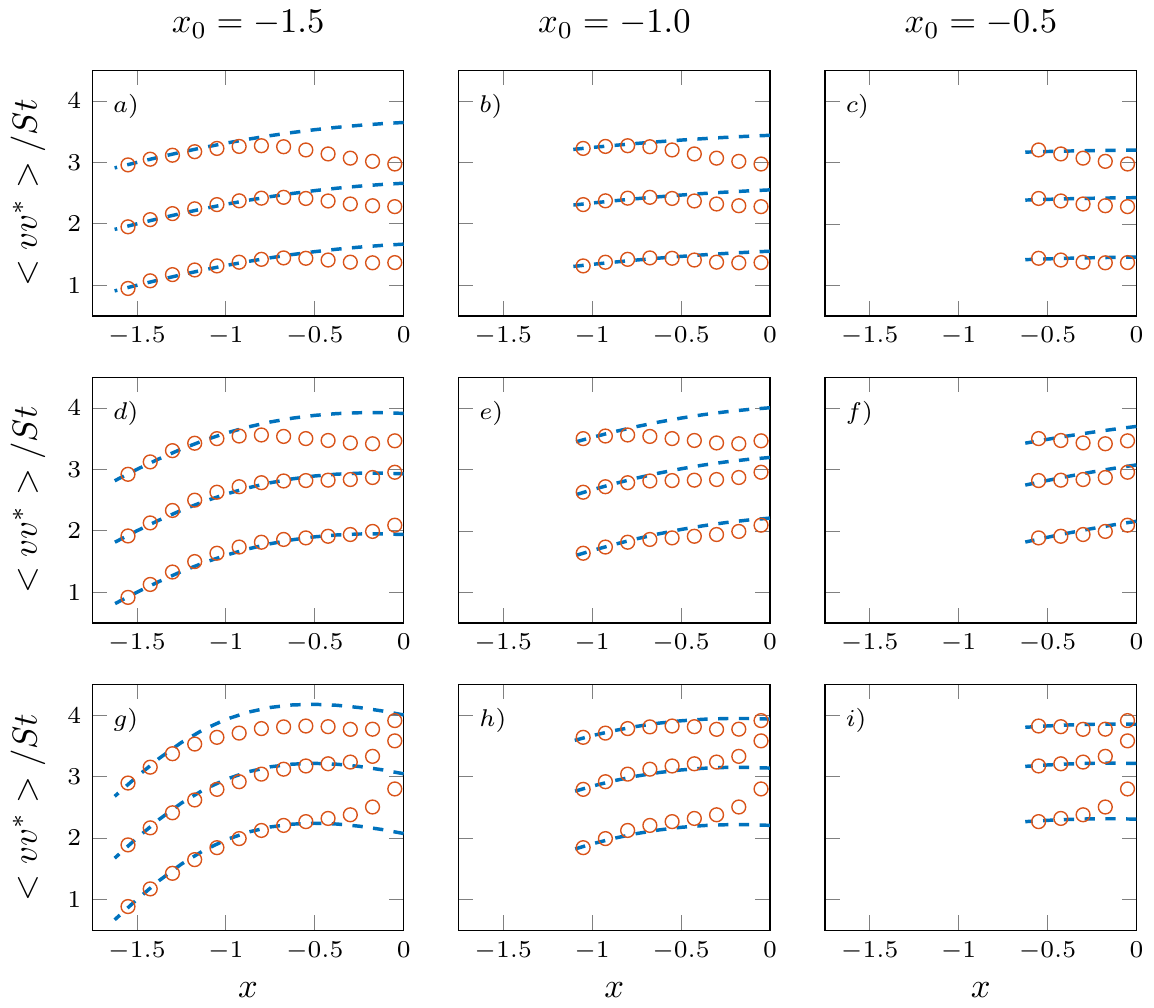}
	\caption{Normalised streamwise envelope of radial velocity fluctutations ($m=0$) at several radial positions for (a,b,c) $St=0.4$ , (d,e,f) $St=0.6$, (g,h,i) $St=0.8$: Leading SPOD mode (red dots), \textit{rank-B} stability model (blue dashed lines). From bottom to top for each subplot: $r=0.38,0.4,0.42$. The curves are shifted by increment of 1 with increasing $r$. Initial conditions of the modal reconstructions are determined from orthogonal projection of the leading SPOD mode onto the stability model at streamwise station $x_0$.}
	\label{fig:decay_disc}
\end{figure}

The results indicate that nozzle structures are essentially underpinned by non-modal dynamics involving the combined action of stable eigenmodes. Such dynamics have been shown to be important also in the downstream region of the jet, and particularly in regions where Kelvin-Helmoltz mode becomes stable \citep{tissot2017wave, towne2018spectral,schmidt2018spectral,lesshafft2019resolvent}. As discussed earlier, the Orr mechanism is responsible for non-modal growth of disturbances in convectively stable turbulent mean flow, and also appears to play a key role in the evolution of perturbations in logarithmic layer of turbulent boundary layers \citep{jimenez2013linear}. One of its defining features is the successive amplification and decay of radial/wall-normal velocity fluctuations, accompanied by streamwise tilting of boundary-layer structures into the mean-flow direction. A comparison is therefore made between the leading SPOD mode and the \textit{rank-B} wavepacket model in terms of streamwise envelopes of radial velocity fluctuations to determine whether this mechanism is active in the nozzle boundary layer. The streamwise evolution of reconstructed perturbation fields, to which each constituent eigenmode contributes with its proper decay rate, is described by equation \eqref{eq:spatial_recons}. Figure \ref{fig:decay_disc} plots the streamwise evolution of radial velocity fluctuations for $m=0$ at several radial positions $r > 0.38$ in the nozzle region. This comparison shows the growth and subsequent decay of the most-energetic nozzle structures towards the jet-exit plane, and that over a range of frequencies, this evolution is reproduced by the \textit{rank-B} stability model containing only eigenmodes from the boundary-layer branch. We here note that the comparisons at $St<0.2$ showed somewhat worse agreement between the model and data; the reason for that may be attributed to relatively poor convergence of the leading SPOD mode at these Strouhal numbers.

%

\begin{figure}
	\centering
	\includegraphics[scale=1]{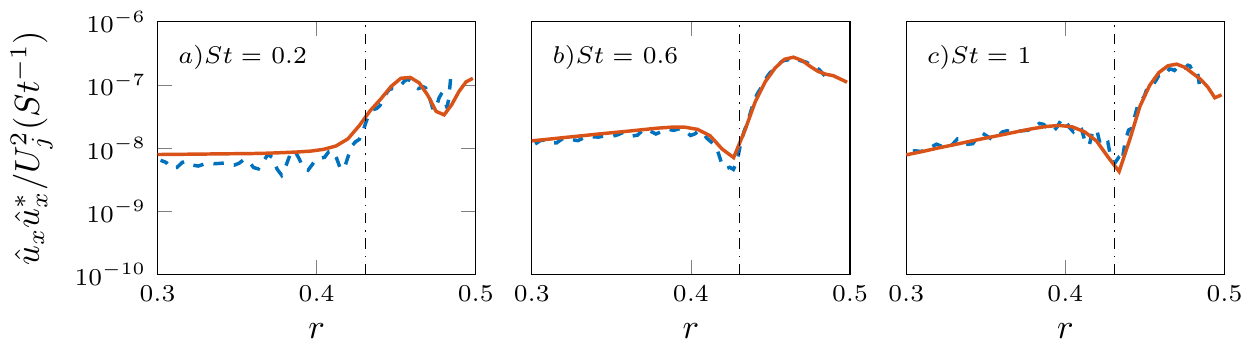}
	\caption{Radial profiles of the streamwise velocity ($m=0$) at $x=-1$: The leading SPOD mode (red solid line) and \textit{rank-BF} stability model (blue dashed line). The model comprises eigenmodes from free-stream branch ($M=60$) and the entire boundary-layer branch.}
	\label{fig:U_EigProjvsPOD_m00_weighted_pipe_x-10_rank_BF}
\end{figure}

\begin{figure}
	\centering
	\includegraphics[scale=1]{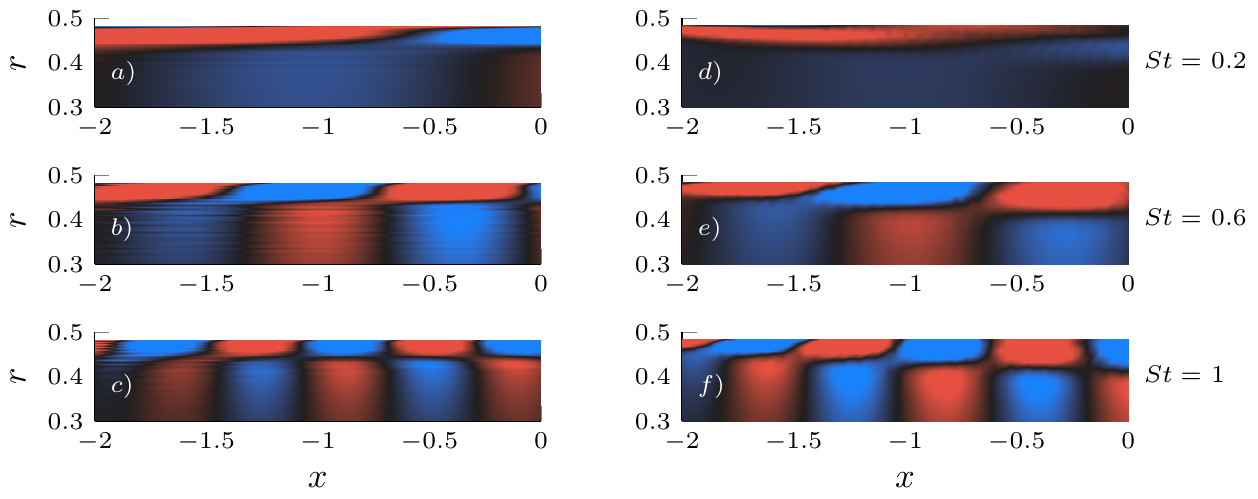}
	\caption{Spatial fields of streamwise velocity fluctuations for $m=0$ within the nozzle: (a,b,c) \textit{rank-BF} stability model; (d,e,f) leading SPOD mode. Real part is shown.}
	\label{fig:Global_U_EigProj_m00_rank_bf}
\end{figure}

Finally, free-stream modes are included in the projection (\textit{rank-BF}). With the contribution of these modes (figures \ref{fig:U_EigProjvsPOD_m00_weighted_pipe_x-10_rank_BF} and \ref{fig:Global_U_EigProj_m00_rank_bf}), the agreement is further improved, as should be expected given that projection with the entire eigenbasis, which is complete, should result in an exact description of the data. However, comparison of the \textit{rank-B} and \textit{rank-BF} reconstructions shows how the essential details of the nozzle boundary-layer structures have been captured by the former. This confirms that the coherent boundary-layer structures, shown earlier to be connected to wavepackets in the jet, may be described with a low-rank, non-modal, stability model.

\section{Resolvent analysis}\label{resolvent_nozzle}

\subsection{Local resolvent analysis}\label{resolvent_nozzle}

Modal decomposition of the leading SPOD modes using the stability eigenbasis has revealed non-modal behaviour, for which the resolvent or input-output framework is most appropriate. This framework is useful for the description of coherent structures in fluid systems with stable eigenvalues, and has been extensively employed in the study of turbulent jets \citep{garnaud2013preferred,schmidt2018spectral,lesshafft2019resolvent,nogueira2019large,cavalieri2019wave,pickering2020lift} and wall-bounded flows \citep{mckeon2010critical,hwang2010linear,abreu2020spod}. The dynamics are conceptually understood as being excited by external forcing that may be associated with ambient noise or non-linear flow interactions.

Resolvent analysis of the turbulent mean nozzle flow is first performed in a locally parallel framework in cylindrical coordinates. The incompressible linearised Navier-Stokes equations, with variables Fourier transformed in the streamwise direction and in time, and expanded in a Fourier series in azimuth, are thus arranged in matrix form as an input-output system

\begin{equation}
(-i\omega{\boldsymbol{I}} - {\boldsymbol{A}}) \left[\begin{array}{c} \hat{u}_x \\ \hat{u}_r \\  \hat{u}_{\theta} \\ \hat{p} \end{array} \right]({\alpha},m,\omega)= \left[\begin{array}{c} \hat{f}_x \\ \hat{f}_r \\  \hat{f}_{\theta} \\ 0 \end{array} \right] ({\alpha},m,\omega),
\end{equation}

\noindent where ${\hat{\boldsymbol{f}}}=[\hat{f}_x \; \hat{f}_r \;  \hat{f}_{\theta} \; 0]^{T}$ denotes momentum forcing associated with non-linear terms; there is no forcing for the continuity equation as it is linear in the incompressible Navier -Stokes equations. The Fourier notation introduced in the preceding linear stability analysis is maintained. Note that streamwise and azimuthal wavenumber, and frequency dependency of the operator $\boldsymbol{A}$ is implicit, and all the related subscripts are dropped in the
interest of simplicity. In addition, matrices $\boldsymbol{B}$ and $\boldsymbol{C}$ are introduced, in order to impose a forcing $\hat{\boldsymbol{f}}$ uniquely to the momentum equations and to obtain an associated response/output $\hat{\boldsymbol{y}}$ consisting of three components of velocity and pressure fluctuations. These matrices are also used to filter out the torsional modes for axisymmetric ($m=0$) disturbances. The input-output system can be written with operators in discrete and compact form as 

\begin{gather} \label{eq:input_output}
(-i\omega{\boldsymbol{I}} - {\boldsymbol{A}}){\hat{\boldsymbol{q}}}=\boldsymbol{{{L}}}{\hat{\boldsymbol{q}}}=\boldsymbol{B}\hat{\boldsymbol{f}}, \\
\hat{\boldsymbol{y}}=\boldsymbol{{C}}{\hat{\boldsymbol{q}}},
\end{gather}

\noindent where $\boldsymbol{{{L}}}$ is the linearised Navier-stokes operator, $\hat{\boldsymbol{q}}$ the state vector of flow variables, and $\hat{\boldsymbol{f}}$ stands for the unknown input/forcing. The matrix operators are provided in \S \ref{app_operators_resolvent}. The relation between input and output can be established by introducing the resolvent operator $\boldsymbol{{{R}}}$ as

\begin{equation}
\label{e:resolvent}
\hat{\boldsymbol{y}}=\boldsymbol{{C}} \boldsymbol{{{L}}}^{-1} \boldsymbol{{B}}
\hat{\boldsymbol{f}}=\boldsymbol{{{R}}} \hat{\boldsymbol{f}}.
\end{equation}

\noindent The objective of the resolvent analysis is to obtain a forcing of minimum norm that maximises the norm of an associated response. This can be achieved by considering the Rayleigh quotient

\begin{equation}
\label{e:rayleigh_quo}
\mathcal{G}^2_{max}(\hat{\boldsymbol{f}})=  \text{ max } \frac{{<\hat{\boldsymbol{y}},\hat{\boldsymbol{y}}>}_{\boldsymbol{{W}}_y}}{{<\hat{\boldsymbol{f}},\hat{\boldsymbol{f}}>}_{\boldsymbol{{W}}_f}}= \frac{{{||{\boldsymbol{{{R}}} \hat{\boldsymbol{f}}}||}^2}_{\boldsymbol{{W}}_y}}{{{||{\hat{\boldsymbol{f}}}||}^2}_{\boldsymbol{{W}}_f}},
\end{equation}

\noindent where $||.||_{\boldsymbol{{W}}}$ stands for the weighted norm under the induced integration weight $\boldsymbol{{W}}$ derived from Chebyshev polynomials. ${\boldsymbol{{W}}_f}=\boldsymbol{{W}}$ is chosen for the input space. A weighting function, similar to what is used in \citet{nogueira2019large}, is introduced to restrict the resolvent response modes within the confines of the nozzle boundary layer ($r>{r}_{\delta})$ 

\begin{equation}
\label{e:resolvent_weight}
{\boldsymbol{{W}}_y}= 0.5 \bigg[1+\tanh{\bigg(\frac{r- {r}_{\delta}}{{\delta}_{1}}\bigg)} \bigg] \boldsymbol{{W}},
\end{equation}

\noindent wherere $r_{\delta}$ is the radial position of the edge of the boundary layer and ${\delta}_{1}$ the displacement thickness. The weight function constrains the resolvent analysis to characterise the fluctuations in the nozzle boundary layer ($r>{r}_{\delta}$) and removes modes associated with free-stream disturbances \citep{nogueira2020resolvent}. Singular-value decomposition of the weighted resolvent operator allows the determination of forcing-response modes that maximise the Rayleigh quotient

\begin{equation}
\label{e:SVD}
{\boldsymbol{{W}}_y^{1/2}}\boldsymbol{\mathcal{{R}}}{\boldsymbol{{W}}_f^{-1/2}}=\boldsymbol{U}\boldsymbol{{\Sigma}}\boldsymbol{V^{*}},
\end{equation}

\noindent where $(.)^{\ast}$ denotes a conjugate transpose. $\boldsymbol{W_{(\cdot)}^{1/2}}$ are computed through Cholesky decomposition of the weight matrices $\boldsymbol{W_{(\cdot)}}=\boldsymbol{W_{(\cdot)}^{1/2}} {\boldsymbol{W_{(\cdot)}^{1/2}}}^{*}$. $\boldsymbol{U}=\left[U^{(1)},U^{(2)},....,U^{(N)} \right]$ and $\boldsymbol{V}=\left[V^{(1)},V^{(2)},....,V^{(N)}\right]$ are left and right singular vectors, and $\boldsymbol{{\Sigma}}$ is a diagonal matrix containing singular values $\boldsymbol{{\Sigma}}=\diag \left({\sigma}^{(1)},{\sigma}^{(2)},...,{\sigma}^{(N)}\right)$ in decreasing order. Equation \eqref{e:SVD} can be rearranged to define forcing $\boldsymbol{{\Psi}}$ and response $\boldsymbol{{\Phi}}$ bases in terms of left and right singular-vectors of the modified resolvent operator as,

\begin{align}
{\boldsymbol{\Psi}} &=\boldsymbol{W^{-1/2}} \boldsymbol{{U}} \\
{\boldsymbol{\Phi}} &=\boldsymbol{W^{-1/2}} \boldsymbol{V}.
\end{align}

\noindent The response and forcing modes are orthonormal in their respective space

\begin{equation}
\label{e:ortho_resolvent}
{<{{\Psi}}^{(i)},{{\Psi}}^{(j)}>}_{{{W}}_y}=
{<{{\Phi}}^{(i)},{{\Phi}}^{(j)}>}_{\boldsymbol{{W}}_f}=\delta_{ij}.
\end{equation}

\noindent Equation \eqref{e:resolvent}, together with the recently introduced variables, can be cast in the following final form:

\begin{equation}
\hat{\boldsymbol{y}}=\boldsymbol{{R}}\hat{\boldsymbol{f}}=\boldsymbol{{\Psi}}\Sigma\boldsymbol{{\Phi}^{*}\boldsymbol{W}\hat{\boldsymbol{f}}},
\end{equation}

\noindent which shows that if a forcing  $\hat{\boldsymbol{f}}={{\Phi}^{(1)}}$ is applied, the output is response mode amplified by the first singular value, i.e. $\hat{\boldsymbol{y}}=\sigma^{(1)}{{\Psi}^{(1)}}$. This forcing-response pair is accordingly referred to as the optimal resolvent mode; the singular value $\sigma^{(1)}$  indicates the gain associated with the mode pair.

The turbulent mean flow, numerical setup, boundary conditions, Reynolds number and grid parameters of the analysis are identical to those introduced in the preceding linear stability analysis. The governing equations are likewise non-dimensionalised with the local displacement thickness and the local mean centerline velocity. The analysis is performed for the same $m-\omega$ pairs explored in the previous section. The discussion of findings mainly focuses on the axisymmetric component of the flow dynamics, which guides the analysis regarding the other low azimuthal modes of interest.


\begin{figure}
	\centering
	\hspace*{-0.62cm}
	\includegraphics[scale=0.8]{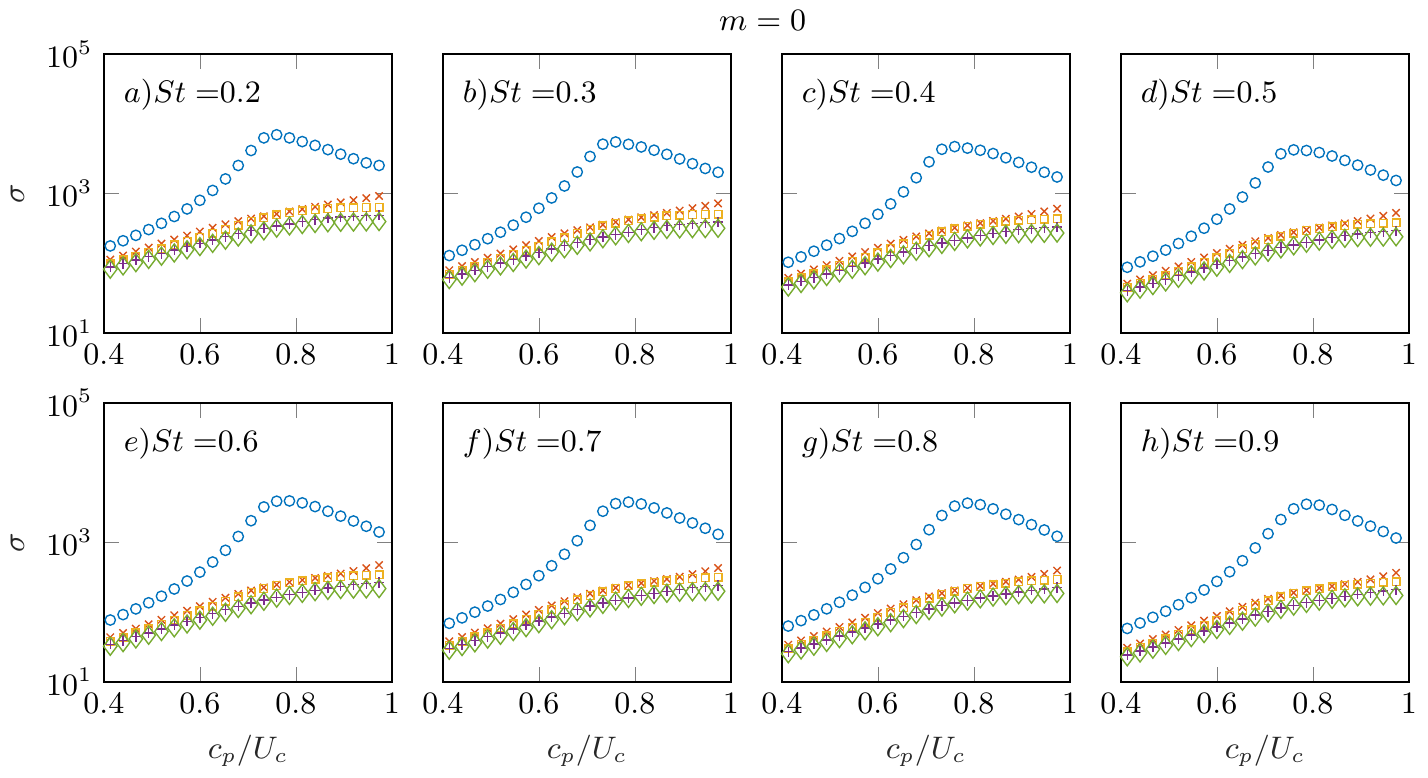}
	\caption{Resolvent gain curves resolved as function of disturbance phase speed $c_p$ for $m=0$ at $x=-1.5$: \textcolor{mycolor1}{\large $\circ$} Optimal mode; \textcolor{mycolor2}{$\times$},\textcolor{mycolor3}{\scriptsize $\square$},\textcolor{mycolor4}{{\scriptsize $+$}},\textcolor{mycolor5}{{$\diamond$}} Suboptimal modes in sequential order.}\label{fig:resolvent_gains_m00_x015}
\end{figure}

The locally parallel formulation of resolvent analysis requires a streamwise wavenumber $\alpha$ to be specified in addition to azimuthal wavenumber and frequency $m-{\omega}$. First, a set of wavenumbers associated with phase speed $c_p={\omega/{\alpha}}$, varying between 0.4 and 1 of the local nozzle mean centerline velocity $U_c$ are considered. This permits the calculation of resolvent gains resolved as a function of phase speed for a given $m-\omega$ pair. Figure \ref{fig:resolvent_gains_m00_x015} shows the energy gain curves of the optimal and first four sub-optimal axisymmetric modes at $x=-1.5$. This shows that the optimal forcing-response mode pair dominates considerably, there being a clear gain seperation between this and sub-optimal modes for all frequencies considered. Morever, a strong peak is observed in a range of $c_p=0.7 \approx 0.8$ at this axial position. Resolvent modes in this range are of interest, as they are likely to be observed in the data. Similar findings are reported in the resolvent analysis of the turbulent boundary layer over an airfoil \citep{abreu2017coherent}.

\begin{figure}
	\centering
	\includegraphics[scale=1]{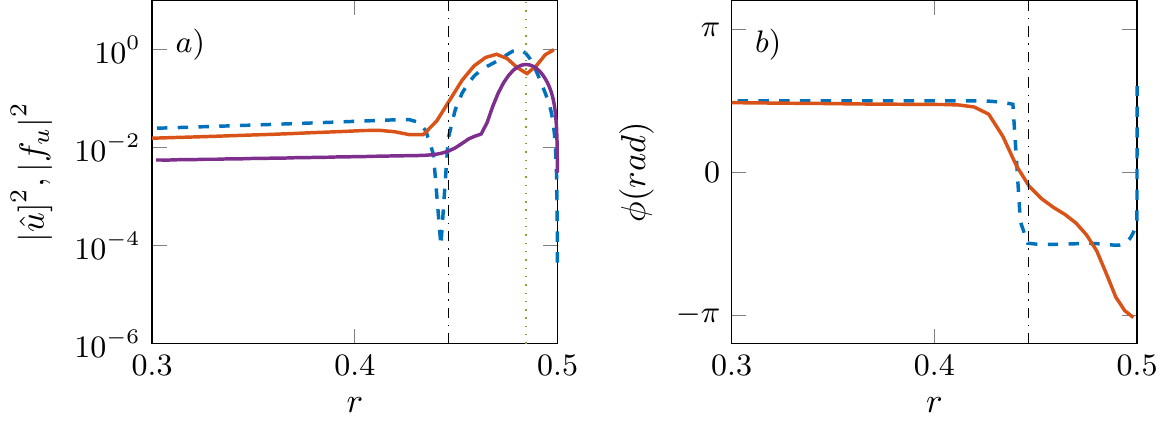}
	\caption{(a) Streamwise velocity profiles for $St=0.4$ at $x=-1.5$. Leading SPOD mode (red solid line); Optimal response mode (blue dashed lines); optimal forcing mode (purple solid line). (b) Phase of streamwise velocity.  Black dash-dot and dotted lines indicate, respectively, local boundary-layer thickness and critical-layer position.}\label{fig:resvspod_St04}
\end{figure}

Figure \ref{fig:resvspod_St04}a shows the radial structure of the axisymmetric ($m=0$) leading SPOD and optimal resolvent modes, all with unit norm, at axial position $x=-1.5$ for frequency $St=0.4$. The resolvent modes are chosen for a phase speed associated with the largest gain $c_{p} \approx0.78$. It is seen that the optimal resolvent response mode is in good agreement with the most energetic nozzle structure educed via SPOD, and that the associated forcing mode is contained inside the nozzle boundary layer. The forcing mode peaks at a radial position associated with the critical-layer, where the local mean velocity is equal to the disturbance phase speed \citep{mckeon2010critical}. It is shown in the previous section that a \textit{rank-1} model based on the homogenous linear problem does not provide a correct description for nozzle wavepackets, which is well captured by a higher-rank truncation of the complete eigenbasis on account of the combination of non-orthogonal eigenmodes, indicating an underlying non-modal mechanism. This non-normality is accounted for in resolvent analysis that implicitly incorporates nonlinearities as an external forcing term, which is here considered spatially white since no model is used for the said non-linear interactions. Resulting optimal resolvent response modes accurately describe the radial structure of the wavepacket educed from SPOD, providing a further evidence for the prevalence of such mechanism in the nozzle boundary layer.

\begin{figure}
	\centering
	\begin{subfigure}{0.48\textwidth}
		\centering
		\includegraphics[trim=0cm 0.0cm 0cm 0.0cm, clip=true,width=1\textwidth]{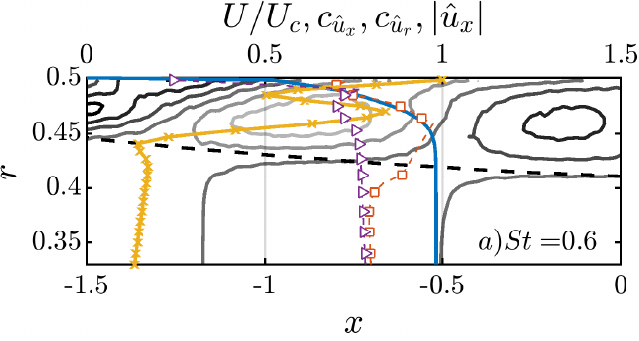}
		\label{fig:phase_St0.6}
	\end{subfigure}
	\begin{subfigure}{0.48\textwidth}
		\centering
		\includegraphics[trim=0cm 0.0cm 0cm 0.0cm, clip=true,width=1\textwidth]{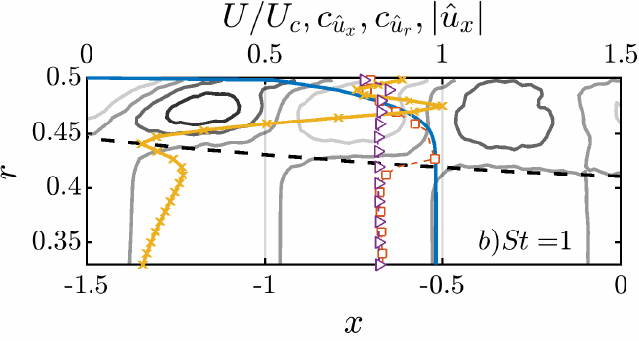}
		\label{fig:phase_St0.8}
	\end{subfigure}
	\caption{Nozzle mean flow (blue solid lines). Contour plots: Real part of streamwise velocity of the leading SPOD mode ($m=0$): a)$St=0.6$; a)$St=0.8$. Radial profiles pertinent to this mode at axial station $x=-1.5$: Phase speed of streamwise velocity (orange square-dashed lines); phase speed of radial velocity (purple triangle-dashed lines); normalised amplitude of streamwise velocity fluctuations (yellow cross-solid lines).}\label{fig:phase_speedNW_m00}
\end{figure}

Figure \ref{fig:resvspod_St04}b also shows that the phase of the leading SPOD mode has a dependence on radial position within the nozzle boundary layer: this is not observed for the optimal response mode, which is likely due to the consideration of a single streamwise wavenumber, and thus of a single phase speed, in the calculation of the response mode. To account for this difference, the phase speed of the leading SPOD mode is computed for two representative frequencies as $c_p \approx\omega({\partial{\phi}/{\partial{x}}})^{-1}$. Figure \ref{fig:phase_speedNW_m00} shows how the streamwise velocity perturbations described by the leading SPOD mode have phase speed that varies across the boundary layer following the trend of the mean velocity profile, indicating that the leading SPOD mode is made up of modes with different wavelengths. Moreover the local analysis precludes inherently the streamwise growth of the boundary layer and associated divergence of the mean flow inside the nozzle. Hence we complement this analysis with a global resolvent approach in which the aforesaid non-parallel mean flow effects are taken into account to obtain a more complete description for the observed low-energy nozzle structures.

\subsection{Global resolvent analysis}


We now move on to characterise and describe the nozzle wavepackets through a global resolvent analysis. Similar to the previous section, we again study linearised Navier-stokes equations recast as an input-output system, while revoking the homogeneity assumption in the streamwise direction and taking the slowly-diverging nature of the mean nozzle flow into consideration. The input-output relationship between nonlinear forcing terms and associated flow responses can be written as:

\begin{subequations}\label{eq:input_output_global}
	\begin{equation} 
(-i\omega{\boldsymbol{I}} - {\boldsymbol{A}}){\hat{\boldsymbol{q}}}=\boldsymbol{{{L}}}{\hat{\boldsymbol{q}}}=\boldsymbol{B}\hat{\boldsymbol{f}}, 
	\end{equation}
		\begin{equation} 
\hat{\boldsymbol{y}}=\boldsymbol{{C}}{\hat{\boldsymbol{q}}},
	\end{equation}
\end{subequations}

\noindent where we maintain the notation introduced in eq. \ref{eq:input_output} and again seek most amplified linear responses to an associated forcing, this time considering a base flow that depends on $x$ and $r$. The linear operators in the global analysis relates response modes solely in frequency - azimuthal wavenumber space and remove the streamwise wavenumber dependence imposed by the local analysis. Starting from the direct input-output system in eq. \ref{eq:input_output_global}, an adjoint input-output system is given by

\begin{subequations}
	\begin{equation}
	\boldsymbol{L}^\dagger \boldsymbol{q}^\dagger = \boldsymbol{B}^\dagger \boldsymbol{f}^\dagger,
	\end{equation}
	\begin{equation}
	\boldsymbol{y}^\dagger = \boldsymbol{C}^\dagger \boldsymbol{q}^\dagger,
	\end{equation}
\end{subequations}
with adjoint operators given in discrete form by $\boldsymbol{L}^\dagger = \boldsymbol{L}^H$, $\boldsymbol{B}^\dagger = \boldsymbol{C}^H\boldsymbol{W}_q$ and $\boldsymbol{C}^\dagger = \boldsymbol{W}_f^{-1}\boldsymbol{B}^H$. An adjoint resolvent operator is thus defined as

\begin{equation}
\boldsymbol{R}^\dagger = \boldsymbol{C}^\dagger(\boldsymbol{L}^\dagger)^{-1}\boldsymbol{B}^\dagger,
\end{equation}
such that forcing modes may be obtained as eigenfunctions of the eigenvalue problem

\begin{equation}
\boldsymbol{R}^\dagger\boldsymbol{R}\boldsymbol{f} = \sigma^2 \boldsymbol{f}.
\end{equation}

For a global analysis, with two-dimensional base-flow and eigenfunctions, this is a large eigenvalue problem more conveniently solved with iterative methods. We have applied the Arnoldi algorithm described by \cite{martini2020efficient}, but differently from that work we have explicitly constructed sparse matrices, with $x$ and $r$ derivatives obtained with fourth-order finite differences and Chebyshev polynomials, respectively. Such sparse matrices allow fast solutions of linear systems using a sparse-matrix LU decomposition, and reduce memory requirements  \citep{gennaro2013sparse}. For each iteration in the Arnoldi method, the action of $\boldsymbol{R}^\dagger\boldsymbol{R}$ onto a test forcing $\boldsymbol{f}_i$ is obtained by computing
\begin{subequations}
	\begin{equation}
	\boldsymbol{L} \boldsymbol{q}_i = \boldsymbol{B} \boldsymbol{f}_i,
	\end{equation}
	\begin{equation}
	\boldsymbol{y}_i = \boldsymbol{C} \boldsymbol{q}_i,
	\end{equation}
	\begin{equation}
	\boldsymbol{L}^\dagger \boldsymbol{w}_i = \boldsymbol{B}^\dagger \boldsymbol{y}_i,
	\end{equation}
	\begin{equation}
	\boldsymbol{f}_{i+1} = \boldsymbol{C}^\dagger \boldsymbol{w}_i,
	\end{equation}
\end{subequations}
such that $\boldsymbol{f}_{i+1} = \boldsymbol{R}^\dagger\boldsymbol{R}\boldsymbol{f}_i$. Thus, the inverses of $\boldsymbol{L}$ and $\boldsymbol{L}^\dagger$ are not computed, as successive solutions of the above equations in Arnoldi iterations are obtained in an efficient manner by using the LU decompositions of $\boldsymbol{L}$ and $\boldsymbol{L}^\dagger$.

A computational domain with $-5 \le x \le 4.1$ and $0 \le r \le 0.5$ was used, with a physical domain given by $-3.5 \le x \le 0$ and sponge zones upstream and downstream of this region. The forcing matrix $\boldsymbol{B}$ restricts the spatial support of the forcing to lie within the computational domain, whereas the observation matrix $\boldsymbol{C}$ considers only responses within the domain $-2 \le x \le 0$ used in the SPOD analysis. A discretisation in 200 points in streamwise and 100 points in radial direction was used. The Arnoldi method was set up using 50 iterations, which was sufficient to converge the leading modes. Changes in the numerical parameters led to negligible differences in the results.

A Reynolds number of $3 \cdot 10^4$ was considered, as in the global resolvent analysis of jets by \cite{schmidt2018spectral}. This value was seen in that work to be sufficiently high to attain asymptotic trends in the resolvent analysis, while maintaining at the same time moderate computational cost. Although not shown here, higher values of $Re$ did not lead to significant changes for global resolvent modes inside the nozzle.

\begin{figure}
	\centering
	\includegraphics[scale=1]{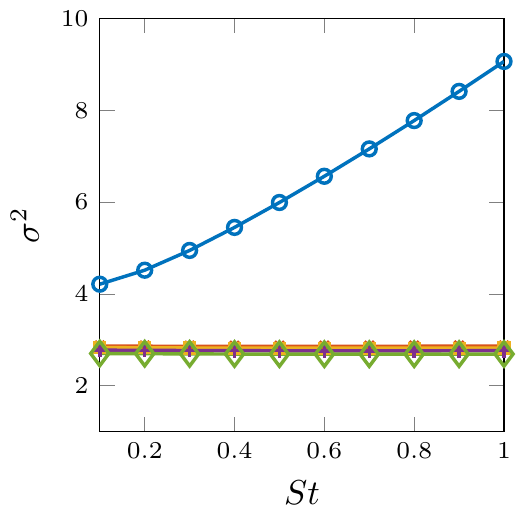}
	\caption{Gain curves of the global resolvent analysis for $m=0$:  \textcolor{mycolor1}{\large $\circ$} Optimal mode; \textcolor{mycolor2}{$\times$},\textcolor{mycolor3}{\scriptsize $\square$},\textcolor{mycolor4}{{\scriptsize $+$}},\textcolor{mycolor5}{{$\diamond$}} Suboptimal modes in sequential order.}
	\label{fig:GR_gains_m00}
\end{figure}

\begin{figure}
	\centering
	\includegraphics[scale=1]{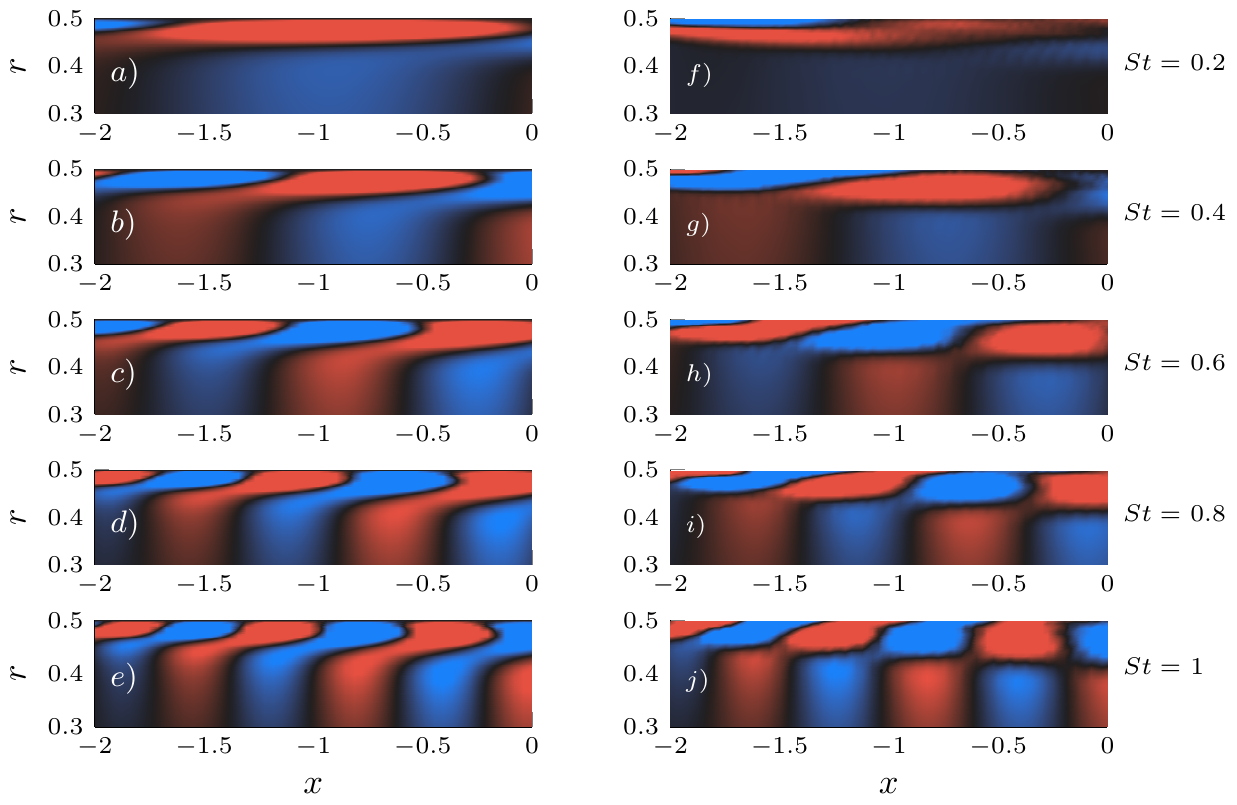}
	\caption{Streamwise velocity of (a-e) optimal resolvent response and  (f-j) SPOD modes within the nozzle ($m=0$) Real part is shown. (${\color{blue}\blacksquare}\!{\color{Black}\blacksquare}\!{\color{red}\blacksquare}$, $\pm$  0.25 of the maximum absolute value in the nozzle)}
	\label{fig:Global_U_ResvsPOD_m00_wp}
\end{figure}

\begin{figure}
	\centering
    \includegraphics[scale=1]{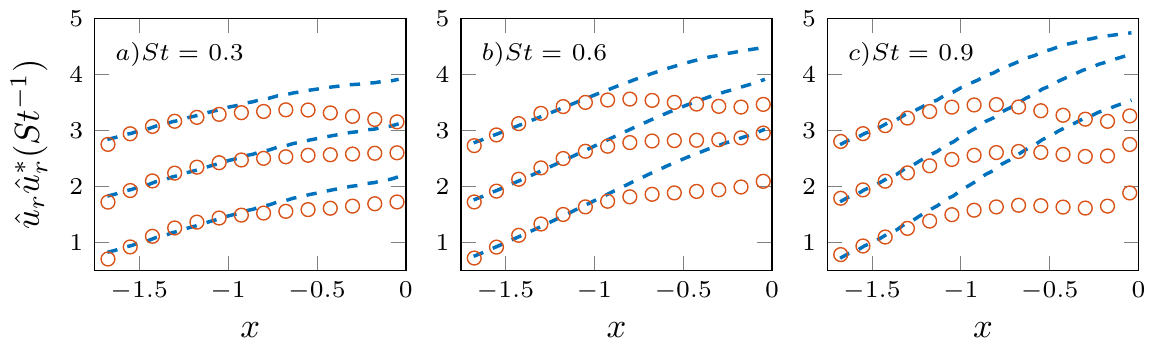}
	\caption{Evolution of radial velocity fluctuations ($m=0$) along several axial slices: leading SPOD (red dots) and optimal response modes (blue dashed lines). From bottom to top for each subplot: $r=0.38,0.4,0.42$. The curves are shifted by increment of 1 with increasing $r$. The modes are normalised by their corresponding amplitudes at $x=-1.5$.}
	\label{fig:decay_res_m00}
\end{figure}

Gain curves, resolved as a function of frequency, for the axisymmetric disturbances are shown in figure \ref{fig:GR_gains_m00}. It is seen that the gain ratio between the optimal and sub-optimal resolvent modes are of same order. This is particularly pronounced at lower $St$: the low-gain seperation suggests that the specific details of the forcing will likely be of significance for the linear flow response mode. We compare global shapes of the leading SPOD and the optimal resolvent modes in figure \ref{fig:Global_U_ResvsPOD_m00_wp}. The figure shows that spatial organisation and essential characteristics of the nozzle wavepackets are reproduced by the resolvent response modes: the resolvent analysis provide axially organised structures that match SPOD modes, and these structures tilt into the mean-flow direction towards the nozzle-exit plane. In the global resolvent analysis the gain ratio between the optimal and sub-optimal modes for low St becomes close to 1 (figure \ref{fig:GR_gains_m00}), and the streamwise support of both resolvent and SPOD modes exceed the nozzle length, which help explaining the discrepancies. Recalling the observation made in the linear stability analysis, the axial envelope of radial velocity fluctuations in SPOD modes have shown an algebraic amplitude growth as the tilting nozzle wavepackets are advected towards the nozzle exit. In figure \ref{fig:decay_res_m00}, we select three frequency - wavenumber pairs from the optimal resolvent response modes and compare their streamwise evolution against those pertaining to the structures observed in the data. It is seen that the optimal response follows the initial trend of amplitude growth of the SPOD modes; the discrepancies further downstream may be attributed to the fact that the resolvent modes are calculated in a fully internal flow, hence the amplitude saturation of SPOD modes associated with nozzle-lip effects are not captured in the resolvent analysis. 

\begin{figure}
	\centering
	\includegraphics[scale=1]{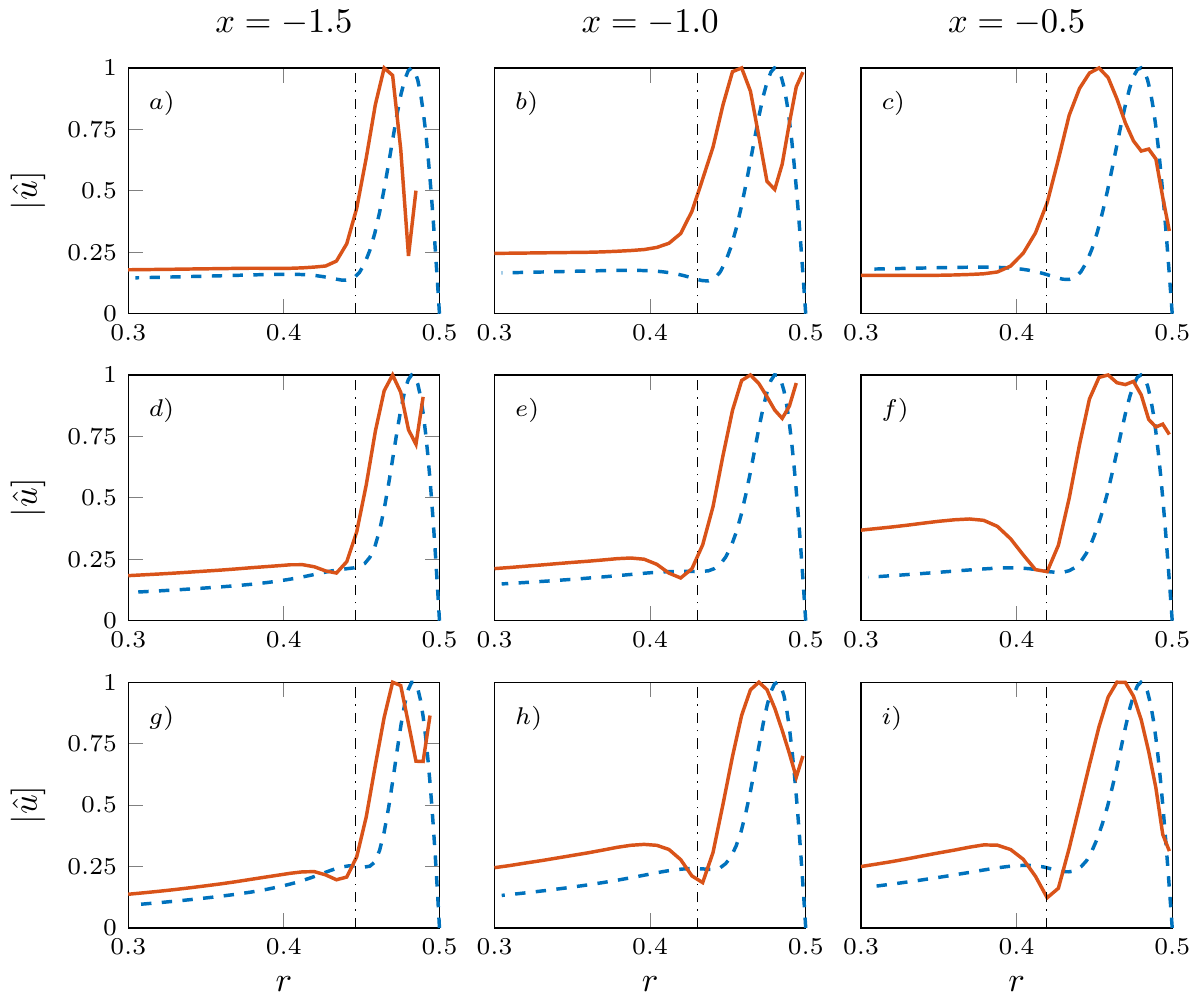}
	\caption{Comparison of streamwise velocity ($m=0$): Leading SPOD (red solid lines) and optimal resolvent response modes (blue dashed lines); $St=0.2$ (a-c),  $St=0.5$ (d-f) , $St=0.8$ (g-i). Dotted line is the local boundary-layer thickness.}
	\label{fig:U_ResProjvsPOD_m00_wp_x-15a}
\end{figure}

\begin{figure}
	\centering
    \includegraphics[scale=1]{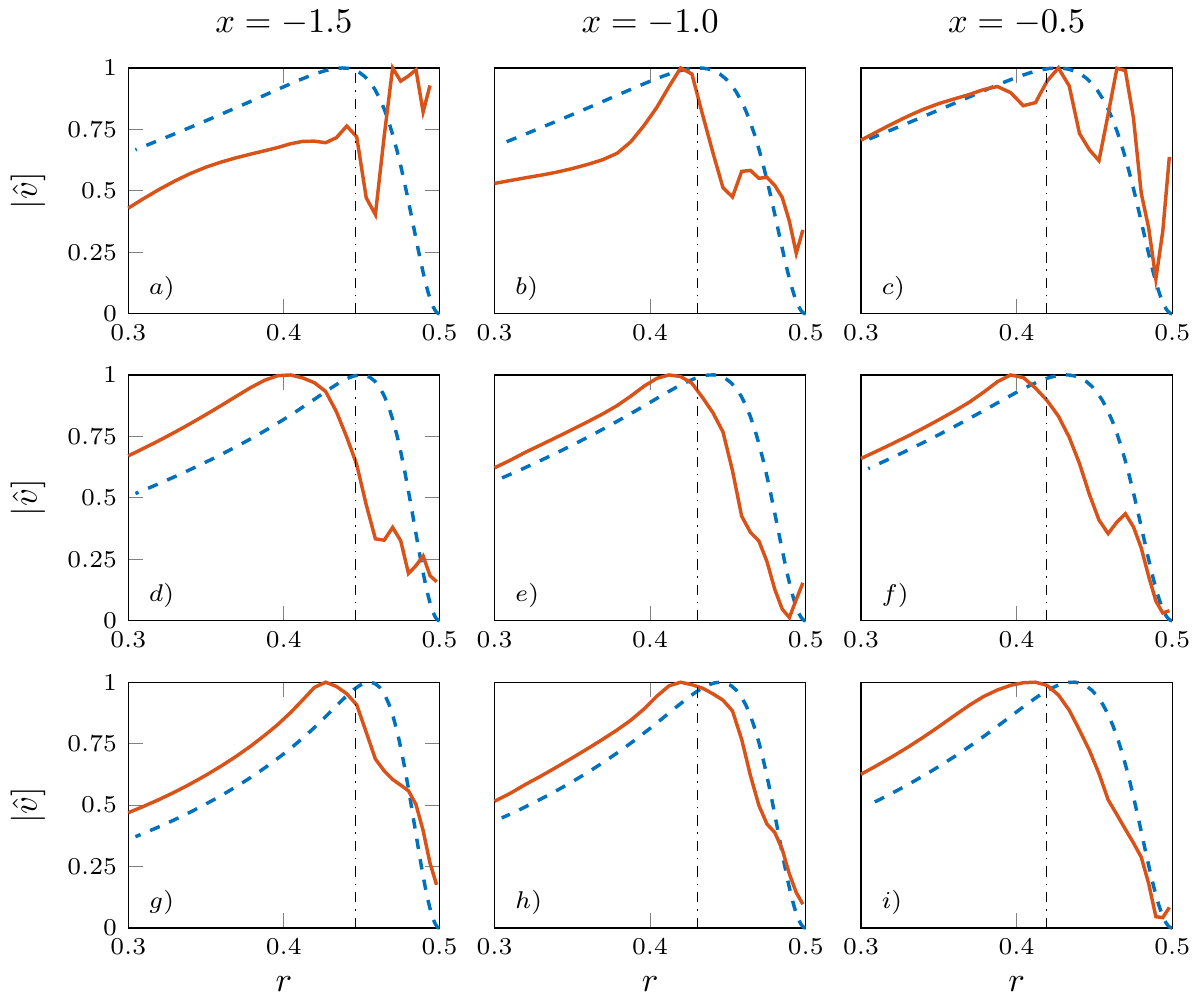}
	\caption{Comparison of radial velocity ($m=0$): Leading SPOD (red solid lines) and optimal resolvent response modes (blue dashed lines); $St=0.2$ (a-c),  $St=0.5$ (d-f) , $St=0.8$ (g-i) . Dotted line is the local boundary-layer thickness.}
	\label{fig:V_ResProjvsPOD_m00_wp_x-15a}
\end{figure}

Figure \ref{fig:U_ResProjvsPOD_m00_wp_x-15a} shows a comparison between the optimal resolvent response and SPOD modes, both obtained inside the nozzle. Over a range of frequencies, radial envelope and support of the nozzle wavepackets are mostly retained by the resolvent modes, which also capture the peak radial position of the SPOD modes for different streamwise positions. A comparison of radial velocity fluctuations similarly show a good agreement (figure \ref{fig:V_ResProjvsPOD_m00_wp_x-15a}), indicating that local characteristics of the nozzle wavepackets are approximated well by the resolvent response modes. 

\begin{figure}
	\centering
    \includegraphics[scale=1]{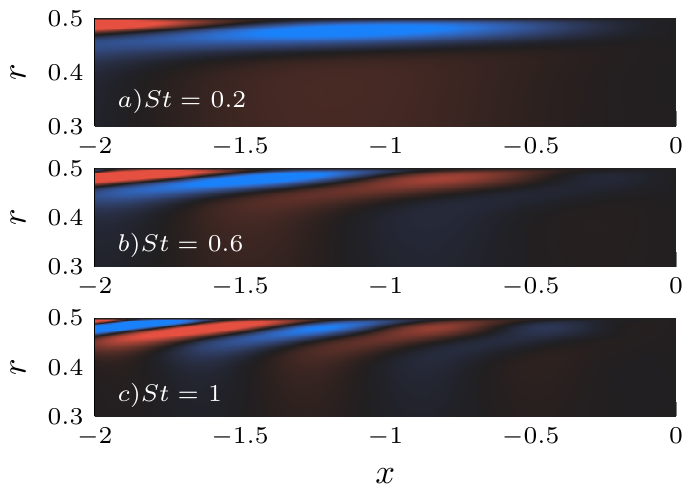}
	\caption{Optimal forcing modes associated with the largest gain for $m=0$. (${\color{blue}\blacksquare}\!{\color{Black}\blacksquare}\!{\color{red}\blacksquare}$, $\pm$  0.25 of the maximum absolute value in the nozzle)}
	\label{fig:resolvent_F_m00a}
\end{figure}

Finally, optimal forcing fields are shown in figure \ref{fig:resolvent_F_m00a}, where a tilting of the forcing modes against the mean shear at angle $45 \degree$, characteristic of the Orr mechanism, is evident. Equally informative is the overlap of the forcing and responses: the response modes are mostly activated near their spatial support (figures \ref{fig:Global_U_ResvsPOD_m00_wp} and \ref{fig:resolvent_F_m00a}). This suggest that a locally parallel analysis may also be an appropriate departure point for modelling nozzle wavepackets, which we have already demonstrated in the previous section. Although such forcing modes are optimal in driving structures inside the nozzle, they are similar to the optimal forcing inside the nozzle that excite downstream wavepackets, as found in global resolvent analysis by \citet{garnaud2013preferred} and \citet{lesshafft2019resolvent}. They showed that the associated Kelvin-Helmholtz instability mechanism is most efficiently activated by the Orr mechanism inside the nozzle. The present analysis with the turbulent nozzle mean-flow confirms that the forcing excites wavy disturbances inside the nozzle, which are advected and trigger the jet wavepackets. 

\section{Conclusions and outlook}

We have considered a subsonic turbulent jet issuing from a cylindrical nozzle at high Reynolds number, by exploiting large-eddy simulation data, to study how fluctuations in the turbulent boundary layer inside the nozzle excite large-scale structures responsible for sound generation in the jet. 

The nozzle mean flow is first characterised in terms of integral boundary-layer quantities, and comparisons with canonical zero-pressure-gradient turbulent boundary layers in the literature confirm the turbulent state of the nozzle flow upstream of the jet exit. Azimuthal decomposition of the streamwise velocity field reveals that low azimuthal wavenumbers contain significantly lower, but non-zero, fluctuation levels with respect to the higher wavenumbers, similar to the azimuthal structure observed downstream of the jet exit \citep{cavalieri2013wavepackets}. 

The fluctuations of low azimuthal wavenumbers in the nozzle boundary layer are shown to have significant cross-spectral density with fluctuations in the region downstream of the jet exit, indicating that two dynamics are connected. SPOD analysis, used to educe wavepackets in both regions, reveals a link between coherent structures in the nozzle and jet regions. The analysis shows that the most-energetic, low-order azimuthal nozzle boundary-layer structures are those that drive the downstream jet dynamics. 

The nozzle wavepackets are characterised using a local linear stability analysis of the turbulent mean flow. It is shown that salient features of the nozzle boundary-layer structures can be described by a low-rank stability model, based on a truncation of the eigenbasis so as to retain a small number of stable boundary-layer eigenmodes. This suggests that the educed structures are underpinned by non-modal dynamics; specifically, the Orr mechanism.

The nozzle boundary-layer structures are further analysed using resolvent analysis, in which non-modal linear dynamics are understood to be driven by non-linear flow interactions treated as an external forcing. The results show that the essential characteristics of the educed nozzle boundary-layer structures can be described by a resolvent analysis, both through local and global approaches. The associated forcing structures are found to tilt against mean-shear in the boundary layer. This confirms the non-modal character of nozzle boundary-layer structures that excite wavepackets downstream of the jet exit. 

The results suggest that reduced-order modelling frameworks for wavepackets should include the flow within the nozzle as the dynamics here likely play an important role in fixing the amplitude of wavepackets in turbulent jets, consistent with other studies that highlight the importance of the nozzle boundary layer
\citep{fontaine2015very,bres2018importance}. The presence of wavepackets as large as nozzle boundary-layer thickness also raises the question of their exploitation for flow and noise control.

\section*{Acknowledgements}

The LES studies were supported by NAVAIR SBIR projects with computational resources provided by DoD HPCMP.

\section*{Declaration of interests}

The authors report no conflict of interest.

\appendix

\section{SPOD of helical modes }\label{SPOD_App}

\begin{figure}
	\centering
	\begin{subfigure}{\textwidth}	
	\centering
	\hspace*{-0.1cm} 
	\includegraphics[scale=1]{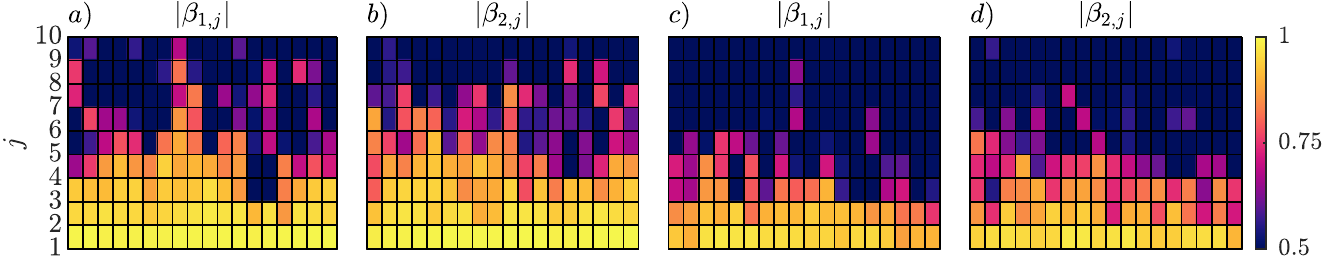}
	
	
	\includegraphics[scale=1]{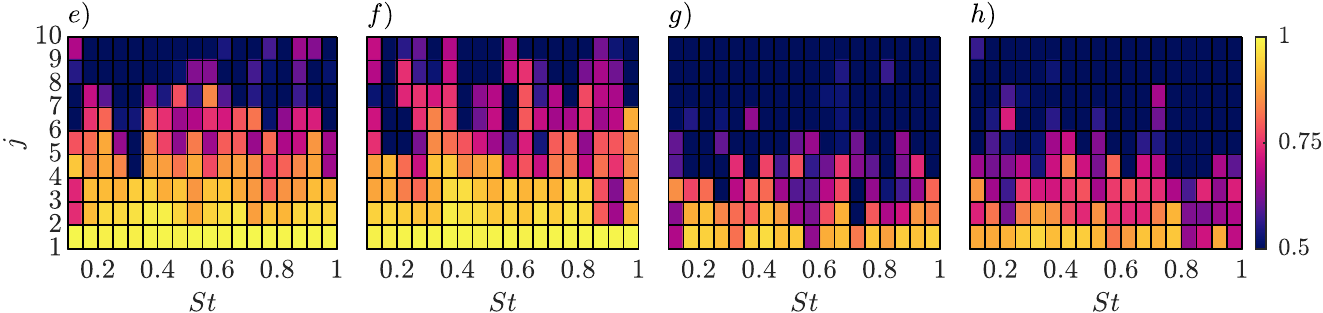}
    \end{subfigure}
	\caption{Convergence maps of SPOD modes for $m=1$ (top row) and $m=2$ (bottom row): (a,b,e,f) `Nozzle-Weighted' SPOD modes; (c,d,g,h) `Jet-Weighted' SPOD modes.}
	\label{fig:SPOD_convergence_m12}
\end{figure}

\begin{figure}
	\centering
	\includegraphics[scale=1]{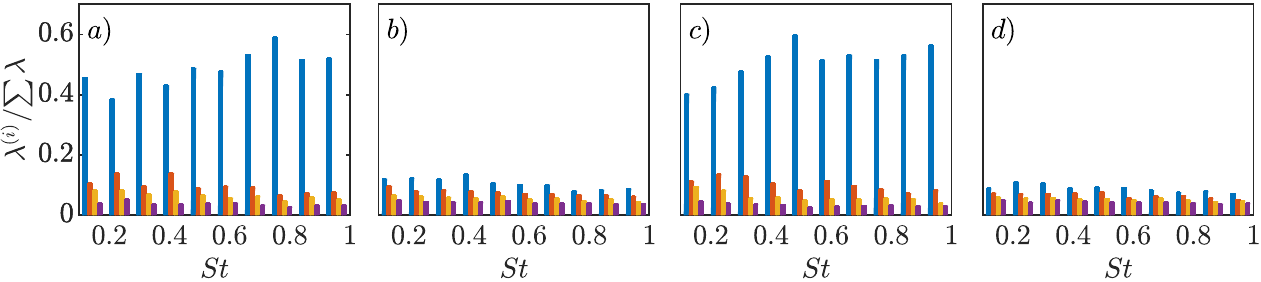}
	\caption{SPOD eigenvalue spectra for `Nozzle-Weighted' (a,c) and `Jet-Weighted' (b,d) as a function of frequency: (a,b) $m=2$; (c,d) $m=3$.}
	\label{fig:SPOD_energies_m23}
\end{figure}

Convergence analyses of SPOD are made for the first two helical modes ($m=1,2$) and figure \ref{fig:SPOD_convergence_m12} shows the associated convergence maps. Similar to the axisymmetric disturbances, the leading modes in both `Nozzle-Weighted' and 'Jet-Weighted' are seen converged for the frequencies we analyse throughout the paper. The energy spectra presented in figure \ref{fig:SPOD_energies_m23} show that the `Nozzle-Weighted' leading mode contains the large fraction of the energy over a range of frequencies, similar to the observations made for the axisymmetric mode.  

\section{Linearised Navier-Stokes equations in cylindrical coordinates}\label{AppC}

\subsection{Mean velocity function} \label{App_VelocityFunction}

A composite velocity function, proposed by \cite{monkewitz2007self}, used to fit the LES mean nozzle flow, which is provided in concise form by \cite{cossu2009optimal} as,
\begin{equation}
U=u_{\tau}\left[U_{i}^{+}\left(y^{+}\right)-U_{\log }^{+}\left(y^{+}\right)+U_{e}^{+}\left({Re}_{{\delta}_{1}}\right)-U_{w}^{+}(\eta)\right]
\end{equation}

$$
U_{w}^{+}(\eta)=\left[\frac{1}{\kappa} E_{1}(\eta)+w_{0}\right] \frac{1}{2}\left[1-\tanh \left(\frac{w_{-1}}{\eta}+w_{2} \eta^{2}+w_{8} \eta^{8}\right)\right]
$$

$$\eta=\frac{y^{+}}{Re_{\delta}}, \ w_{0}=0.6332, \ w_{-1}=-0.096, w_{2}=28.5 \  \text{and}  \ w_{8}=33000,
$$

$$
U_{\log }^{+}\left(y^{+}\right)=\frac{1}{\kappa} \ln \left(y^{+}\right)+B
$$
$$
U_{e}^{+}\left(R e_{\delta .}\right)=\frac{1}{\kappa} \ln \left(R e_{\delta .}\right)+C,
$$
$$
\kappa=0.384, \ B=4.17,\ C=3.3;
$$

$$
U_{i}^{+}\left(y^{+}\right)=0.68285472 \ln \left(y^{+2}+4.7673096 y^{+}+9545.9963\right)
$$
$$
\makebox[2.2cm] \ +1.2408249 \arctan \left(0.010238083 y^{+}+0.024404056\right)
$$
$$
\makebox[1.3cm] \ +1.2384572 \ln \left(y^{+}+95.232690\right)-11.930683
$$
$$
\makebox[2.03cm] \ -0.50435126 \ln \left(y^{+2}-7.8796955 y^{+}+78.389178\right)
$$
$$
\makebox[1.9cm] \ +4.7413546 \arctan \left(0.12612158 y^{+}-0.49689982\right)
$$
$$
\makebox[1.8cm] \ -2.7768771 \ln \left(y^{+2}+16.209175 y^{+}+933.16587\right)
$$
$$
\makebox[2.2cm] \ +0.37625729 \arctan \left(0.033952353 y^{+}+0.27516982\right)
$$
$$
\makebox[1.3cm] \ +6.5624567 \ln \left(y^{+}+13.670520\right)+6.1128254
$$

\subsection{Matrix operators for linearised Navier-Stokes equations} \label{Operators_LST}

The matrix form of the eigenvalue problem considered in the equation \eqref{eq:lst_eig} is given as follows

\begin{equation}\label{spatial_lst_1}
\left[\begin{array}{cccc} 
L_{11} & L_{12} & 0 & 0 \\[2pt]
0 & L_{22} & L_{23} & L_{24} \\[2pt]
0 & L_{32} & L_{33} & L_{34} \\ [2pt]
0 & L_{42}  & L_{43} & L_{44} \\ [2pt]
\end{array} \right]\left[\begin{array}{c}  
\hat{u} \\ [2pt]
\hat{v} \\ [2pt]
\hat{w} \\ [2pt]
\hat{p} \\ [2pt]
\end{array} \right]=\alpha
\left[\begin{array}{cccc} 
F_{11}  & 0   & 0     & F_{14} \\ [2pt]
0     & F_{22} & 0     & 0 \\ [2pt]
0     & 0   & F_{33}   & 0 \\ [2pt]
F_{41} & 0     & 0     & 0 \\ [2pt]
\end{array} \right]
\left[\begin{array}{c} 
\hat{u} \\ [2pt]
\hat{v} \\ [2pt]
\hat{w} \\ [2pt]
\hat{p} \\ [2pt]
\end{array} \right]
\end{equation}

\noindent with $0$ and $I$ as, respectively, zero and identity matrices of $N{\times}N$ dimensions. $N$ is the number of Chebyshev collocation points. $D$ denotes the first order derivative in the radial direction. $m$ and $\omega$ are azimuthal wavenumber and angular frequency respectively, and $Re$ is Reynolds number.

\begin{center}
	\begin{equation}
	\Delta = \quad -\dfrac{m^{2}}{r^{2}} +  \dfrac{1}{r}\dfrac{\partial}{\partial r} + \dfrac{\partial^{2}}{\partial r^{2}}
	\end{equation}
	
	\begin{equation} 
	C =  \dfrac{2 i m}{r^{2}}
	\end{equation}
	\begin{equation}
	L_{11} = -i\omega I - \dfrac{\Delta}{Re} , \qquad  L_{12} = \dfrac{\partial u}{\partial r}
	\end{equation}
	\begin{equation}
	L_{33} = L_{22} = -i\omega I  - \left(\Delta - \dfrac{1}{r}\right)\dfrac{1}{Re} , \qquad L_{23}  = \dfrac{C}{Re}
	\end{equation}
	\begin{equation} 
	L_{24} = \dfrac{\partial}{\partial r}, \qquad L_{32} = \dfrac{-C}{Re}
	\end{equation}
	\begin{equation}
	L_{34} = L_{43} = \dfrac{i m}{r}, \qquad L_{42} = \dfrac{1}{r} + \dfrac{\partial}{\partial r}
	\end{equation}
\end{center}

\begin{equation}
F_{11} = F_{22}= F_{33} = -iU \qquad F_{14}=F_{41}=-iI
\end{equation}

\subsection{Matrix operators of the resolvent analysis} \label{app_operators_resolvent}

The operators of input-output problem (equation \eqref{eq:input_output}) is based on the formulation of \citet{mckeon2010critical} and are provided as follows, 

\begin{gather} \label{eq:input_output_app}
(-i\omega{\boldsymbol{I}} - {\boldsymbol{A}}){\boldsymbol{\hat{q}}}=\boldsymbol{{L}}{\boldsymbol{\hat{q}}}=\boldsymbol{B}\boldsymbol{\hat{f}} \\[10pt]
\boldsymbol{{C}}{\boldsymbol{\hat{q}}} =\boldsymbol{\hat{y}}
\end{gather}

\renewcommand{\arraystretch}{2.5}

\begin{equation}\label{eq_res}
A=\left[\begin{array}{cccc} 
-i{\alpha}_{x}U+\dfrac{\Delta}{Re} & - \dfrac{{\partial U}}{{\partial} r} & 0 & -i{\alpha}_{x}\\
0 & -i{\alpha}_{x}U+\dfrac{1}{Re}\bigg({\Delta-\dfrac{1}{r^2}}\bigg) & \dfrac{E}{Re} & - {\dfrac{\partial}{\partial r}}\\
0 & -\dfrac{E}{Re} & -i{\alpha}_{x}U+\dfrac{1}{Re}\bigg({\Delta-\dfrac{1}{r^2}}\bigg) & -\dfrac{im}{r} \\
i{\alpha}_{x} & \dfrac{1}{r} + \dfrac{{\partial}}{{\partial} r} & \dfrac{im}{r} & 0 \\
\end{array} \right] 
\end{equation}

\begin{equation}
\Delta=-{{\alpha}_{x}}^2-\dfrac{m^2}{r^2}+\dfrac{{\partial}}{{\partial} r} \dfrac{1}{r}+\dfrac{{\partial}^2}{{\partial} r^2}
\end{equation}

\begin{equation}
E=\dfrac{-2im}{r^2}
\end{equation}

with $0$ and $I$ as, respectively, zero and identity matrices of $N{\times}N$ dimensions.

\renewcommand{\arraystretch}{1}

\begin{equation}
\boldsymbol{{B}}=\left[\begin{array}{cccc} 
I & 0  & 0 & 0 \\ 0 & I & 0 & 0 \\ 0 & 0 & I & 0 \\ 0 & 0 & 0 & 0
\end{array} \right]   \text{ and } \boldsymbol{{C}}=\left[\begin{array}{cccc} 
I & 0  & 0 & 0 \\ 0 & I & 0 & 0 \\ 0 & 0 & I & 0 \\ 0 & 0 & 0 & I \\
\end{array} \right] \text{ and } \boldsymbol{{I}}=\boldsymbol{{B}}.
\end{equation}

\clearpage

\bibliography{reference}

\begin{thebibliography}{77}
\expandafter\ifx\csname natexlab\endcsname\relax\def\natexlab#1{#1}\fi
\def\au#1{#1} \def\ed#1{#1} \def\yr#1{#1}\def\at#1{#1}\def\jt#1{\textit{#1}}
  \def\bt#1{#1}\def\bvol#1{\textbf{#1}} \def\vol#1{#1} \def\pg#1{#1}
  \def\publ#1{#1}\def\arxiv#1{#1}\def\org#1{#1}\def\st#1{\textit{#1}}

\bibitem[Abreu {\em et~al.\/}(2020)Abreu, Cavalieri, Schlatter, Vinuesa \&
  Henningson]{abreu2020spod}
{\sc \au{Abreu, L. }, \au{Cavalieri, A. }, \au{Schlatter, P. }, \au{Vinuesa, R.
  } \& \au{Henningson, D. }} \yr{2020}  \at{Spod and resolvent analysis of
  near-wall coherent structures in turbulent pipe flows}.  \jt{arXiv preprint
  arXiv:2004.11452} .

\bibitem[Abreu {\em et~al.\/}(2017)Abreu, Cavalieri \& Wolf]{abreu2017coherent}
{\sc \au{Abreu, L.~I. }, \au{Cavalieri, A.~V. } \& \au{Wolf, W. }} \yr{2017}
  Coherent hydrodynamic waves and trailing-edge noise.  \bt{In {\em 23rd
  AIAA/CEAS Aeroacoustics Conference\/}},  \pg{p. 3173}.

\bibitem[Alizard \& Robinet(2011)]{alizard2011modeling}
{\sc \au{Alizard, F. } \& \au{Robinet, J.-C. }} \yr{2011}  \at{Modeling of
  optimal perturbations in flat plate boundary layer using global modes:
  benefits and limits}.  \jt{Theoretical and Computational Fluid Dynamics}
  \bvol{25}~(1-4),  \pg{147--165}.

\bibitem[Batt(1975)]{batt1975layer}
{\sc \au{Batt, R.~G. }} \yr{1975}  \at{Layer some measurements on the effect of
  tripping the two-dimensional shear}.  \jt{AIAA Journal}  \bvol{13}~(2),
  \pg{245--247}.

\bibitem[Beneddine {\em et~al.\/}(2016)Beneddine, Sipp, Arnault, Dandois \&
  Lesshafft]{beneddine2016conditions}
{\sc \au{Beneddine, S. }, \au{Sipp, D. }, \au{Arnault, A. }, \au{Dandois, J. }
  \& \au{Lesshafft, L. }} \yr{2016}  \at{Conditions for validity of mean flow
  stability analysis}.  \jt{Journal of Fluid Mechanics}  \bvol{798},
  \pg{485--504}.

\bibitem[B{\"o}berg \& Brosa(1988)]{boberg1988onset}
{\sc \au{B{\"o}berg, L. } \& \au{Brosa, U. }} \yr{1988}  \at{Onset of
  turbulence in a pipe}.  \jt{Zeitschrift f{\"u}r Naturforschung A}
  \bvol{43}~(8-9),  \pg{697--726}.

\bibitem[Bogey \& Bailly(2010)]{bogey2010influence}
{\sc \au{Bogey, C. } \& \au{Bailly, C. }} \yr{2010}  \at{Influence of
  nozzle-exit boundary-layer conditions on the flow and acoustic fields of
  initially laminar jets}.  \jt{Journal of Fluid Mechanics}  \bvol{663},
  \pg{507--538}.

\bibitem[Bradshaw(1966)]{bradshaw1966effect}
{\sc \au{Bradshaw, P. }} \yr{1966}  \at{The effect of initial conditions on the
  development of a free shear layer}.  \jt{Journal of Fluid Mechanics}
  \bvol{26}~(2),  \pg{225--236}.

\bibitem[Breakey {\em et~al.\/}(2017)Breakey, Jordan, Cavalieri, Nogueira,
  L{\'e}on, Colonius \& Rodr{\'\i}guez]{breakey2017experimental}
{\sc \au{Breakey, D.~E. }, \au{Jordan, P. }, \au{Cavalieri, A.~V. },
  \au{Nogueira, P.~A. }, \au{L{\'e}on, O. }, \au{Colonius, T. } \&
  \au{Rodr{\'\i}guez, D. }} \yr{2017}  \at{Experimental study of turbulent-jet
  wave packets and their acoustic efficiency}.  \jt{Physical Review Fluids}
  \bvol{2}~(12),  \pg{124601}.

\bibitem[Br{\`e}s {\em et~al.\/}(2017)Br{\`e}s, Ham, Nichols \&
  Lele]{bres2017unstructured}
{\sc \au{Br{\`e}s, G.~A. }, \au{Ham, F.~E. }, \au{Nichols, J.~W. } \& \au{Lele,
  S.~K. }} \yr{2017}  \at{Unstructured large-eddy simulations of supersonic
  jets}.  \jt{AIAA journal}  \pg{pp. 1164--1184}.

\bibitem[Br{\`e}s {\em et~al.\/}(2018)Br{\`e}s, Jordan, Jaunet, Le~Rallic,
  Cavalieri, Towne, Lele, Colonius \& Schmidt]{bres2018importance}
{\sc \au{Br{\`e}s, G.~A. }, \au{Jordan, P. }, \au{Jaunet, V. }, \au{Le~Rallic,
  M. }, \au{Cavalieri, A.~V. }, \au{Towne, A. }, \au{Lele, S.~K. },
  \au{Colonius, T. } \& \au{Schmidt, O.~T. }} \yr{2018}  \at{Importance of the
  nozzle-exit boundary-layer state in subsonic turbulent jets}.  \jt{Journal of
  Fluid Mechanics}  \bvol{851},  \pg{83--124}.

\bibitem[Br{\`e}s \& Lele(2019)]{bres2019modelling}
{\sc \au{Br{\`e}s, G.~A. } \& \au{Lele, S.~K. }} \yr{2019}  \at{Modelling of
  jet noise: a perspective from large-eddy simulations}.  \jt{Philosophical
  Transactions of the Royal Society A}  \bvol{377}~(2159),  \pg{20190081}.

\bibitem[Bridges \& Hussain(1987)]{bridges1987roles}
{\sc \au{Bridges, J. } \& \au{Hussain, A. }} \yr{1987}  \at{Roles of initial
  condition and vortex pairing in jet noise}.  \jt{Journal of Sound and
  Vibration}  \bvol{117}~(2),  \pg{289--311}.

\bibitem[Cavalieri {\em et~al.\/}(2019)Cavalieri, Jordan \&
  Lesshafft]{cavalieri2019wave}
{\sc \au{Cavalieri, A.~V. }, \au{Jordan, P. } \& \au{Lesshafft, L. }} \yr{2019}
   \at{Wave-packet models for jet dynamics and sound radiation}.  \jt{Applied
  Mechanics Reviews}  \bvol{71}~(2),  \pg{020802}.

\bibitem[Cavalieri {\em et~al.\/}(2013)Cavalieri, Rodr{\'\i}guez, Jordan,
  Colonius \& Gervais]{cavalieri2013wavepackets}
{\sc \au{Cavalieri, A.~V. }, \au{Rodr{\'\i}guez, D. }, \au{Jordan, P. },
  \au{Colonius, T. } \& \au{Gervais, Y. }} \yr{2013}  \at{Wavepackets in the
  velocity field of turbulent jets}.  \jt{Journal of fluid mechanics}
  \bvol{730},  \pg{559--592}.

\bibitem[Chu(1965)]{chu1965energy}
{\sc \au{Chu, B.-T. }} \yr{1965}  \at{On the energy transfer to small
  disturbances in fluid flow (part i)}.  \jt{Acta Mechanica}  \bvol{1}~(3),
  \pg{215--234}.

\bibitem[Citriniti \& George(2000)]{citriniti2000reconstruction}
{\sc \au{Citriniti, J. } \& \au{George, W.~K. }} \yr{2000}  \at{Reconstruction
  of the global velocity field in the axisymmetric mixing layer utilizing the
  proper orthogonal decomposition}.  \jt{Journal of Fluid Mechanics}
  \bvol{418},  \pg{137--166}.

\bibitem[Cossu {\em et~al.\/}(2009)Cossu, Pujals \& Depardon]{cossu2009optimal}
{\sc \au{Cossu, C. }, \au{Pujals, G. } \& \au{Depardon, S. }} \yr{2009}
  \at{Optimal transient growth and very large--scale structures in turbulent
  boundary layers}.  \jt{Journal of Fluid Mechanics}  \bvol{619},  \pg{79--94}.

\bibitem[Crighton \& Gaster(1976)]{crighton1976stability}
{\sc \au{Crighton, D. } \& \au{Gaster, M. }} \yr{1976}  \at{Stability of slowly
  diverging jet flow}.  \jt{Journal of Fluid Mechanics}  \bvol{77}~(2),
  \pg{397--413}.

\bibitem[Crow \& Champagne(1971)]{crow1971orderly}
{\sc \au{Crow, S.~C. } \& \au{Champagne, F. }} \yr{1971}  \at{Orderly structure
  in jet turbulence}.  \jt{Journal of Fluid Mechanics}  \bvol{48}~(3),
  \pg{547--591}.

\bibitem[Eitel-Amor {\em et~al.\/}(2014)Eitel-Amor, {\"O}rl{\"u} \&
  Schlatter]{eitel2014simulation}
{\sc \au{Eitel-Amor, G. }, \au{{\"O}rl{\"u}, R. } \& \au{Schlatter, P. }}
  \yr{2014}  \at{Simulation and validation of a spatially evolving turbulent
  boundary layer up to re$\theta$= 8300}.  \jt{International Journal of Heat
  and Fluid Flow}  \bvol{47},  \pg{57--69}.

\bibitem[Fontaine {\em et~al.\/}(2015)Fontaine, Elliott, Austin \&
  Freund]{fontaine2015very}
{\sc \au{Fontaine, R.~A. }, \au{Elliott, G.~S. }, \au{Austin, J.~M. } \&
  \au{Freund, J.~B. }} \yr{2015}  \at{Very near-nozzle shear-layer turbulence
  and jet noise}.  \jt{Journal of Fluid Mechanics}  \bvol{770},  \pg{27--51}.

\bibitem[Freund \& Colonius(2009)]{freund2009turbulence}
{\sc \au{Freund, J. } \& \au{Colonius, T. }} \yr{2009}  \at{Turbulence and
  sound-field pod analysis of a turbulent jet}.  \jt{International Journal of
  Aeroacoustics}  \bvol{8}~(4),  \pg{337--354}.

\bibitem[Garnaud {\em et~al.\/}(2013)Garnaud, Lesshafft, Schmid \&
  Huerre]{garnaud2013preferred}
{\sc \au{Garnaud, X. }, \au{Lesshafft, L. }, \au{Schmid, P. } \& \au{Huerre, P.
  }} \yr{2013}  \at{The preferred mode of incompressible jets: linear frequency
  response analysis}.  \jt{Journal of Fluid Mechanics}  \bvol{716},
  \pg{189--202}.

\bibitem[Gennaro {\em et~al.\/}(2013)Gennaro, Rodr{\'\i}guez, Medeiros \&
  Theofilis]{gennaro2013sparse}
{\sc \au{Gennaro, E. }, \au{Rodr{\'\i}guez, D. }, \au{Medeiros, M. } \&
  \au{Theofilis, V. }} \yr{2013}  \at{Sparse techniques in global flow
  instability with application to compressible leading-edge flow}.  \jt{AIAA
  journal}  \bvol{51}~(9),  \pg{2295--2303}.

\bibitem[Grosch \& Salwen(1978)]{grosch1978continuous}
{\sc \au{Grosch, C.~E. } \& \au{Salwen, H. }} \yr{1978}  \at{The continuous
  spectrum of the orr-sommerfeld equation. part 1. the spectrum and the
  eigenfunctions}.  \jt{Journal of Fluid Mechanics}  \bvol{87}~(1),
  \pg{33--54}.

\bibitem[Grosche(1974)]{grosche1974distributions}
{\sc \au{Grosche, F. }} \yr{1974} Distributions of sound source intensities in
  subsonic and supersonic jets.  \bt{In {\em AGARD, Conf. Proc., 1974\/}}, ,
  \vol{vol. 131},  \pg{pp. 4--1}.

\bibitem[Gudmundsson \& Colonius(2011)]{gudmundsson2011instability}
{\sc \au{Gudmundsson, K. } \& \au{Colonius, T. }} \yr{2011}  \at{Instability
  wave models for the near-field fluctuations of turbulent jets}.  \jt{Journal
  of Fluid Mechanics}  \bvol{689},  \pg{97--128}.

\bibitem[Hanifi {\em et~al.\/}(1996)Hanifi, Schmid \&
  Henningson]{hanifi1996transient}
{\sc \au{Hanifi, A. }, \au{Schmid, P.~J. } \& \au{Henningson, D.~S. }}
  \yr{1996}  \at{Transient growth in compressible boundary layer flow}.
  \jt{Physics of Fluids}  \bvol{8}~(3),  \pg{826--837}.

\bibitem[Hill~Jr {\em et~al.\/}(1976)Hill~Jr, Jenkins \&
  Gilbert]{hill1976effects}
{\sc \au{Hill~Jr, W.~G. }, \au{Jenkins, R.~C. } \& \au{Gilbert, B.~L. }}
  \yr{1976}  \at{Effects of the initial boundary-layer state on turbulent jet
  mixing}.  \jt{AIAA journal}  \bvol{14}~(11),  \pg{1513--1514}.

\bibitem[Hussain \& Zedan(1978{\natexlab{{\em a\/}}})]{hussain1978effects}
{\sc \au{Hussain, A. } \& \au{Zedan, M. }} \yr{1978{\natexlab{{\em a\/}}}}
  \at{Effects of the initial condition on the axisymmetric free shear layer:
  Effect of the initial fluctuation level}.  \jt{The Physics of Fluids}
  \bvol{21}~(9),  \pg{1475--1481}.

\bibitem[Hussain \& Zedan(1978{\natexlab{{\em b\/}}})]{hussain1978effects_mom}
{\sc \au{Hussain, A. } \& \au{Zedan, M. }} \yr{1978{\natexlab{{\em b\/}}}}
  \at{Effects of the initial condition on the axisymmetric free shear layer:
  Effects of the initial momentum thickness}.  \jt{The Physics of Fluids}
  \bvol{21}~(7),  \pg{1100--1112}.

\bibitem[Hwang \& Cossu(2010)]{hwang2010linear}
{\sc \au{Hwang, Y. } \& \au{Cossu, C. }} \yr{2010}  \at{Linear non-normal
  energy amplification of harmonic and stochastic forcing in the turbulent
  channel flow}.  \jt{Journal of Fluid Mechanics}  \bvol{664},  \pg{51--73}.

\bibitem[Jim{\'e}nez(2013)]{jimenez2013linear}
{\sc \au{Jim{\'e}nez, J. }} \yr{2013}  \at{How linear is wall-bounded
  turbulence?}  \jt{Physics of Fluids}  \bvol{25}~(11),  \pg{110814}.

\bibitem[Jordan \& Colonius(2013)]{jordan2013wave}
{\sc \au{Jordan, P. } \& \au{Colonius, T. }} \yr{2013}  \at{Wave packets and
  turbulent jet noise}.  \jt{Annual review of fluid mechanics}  \bvol{45},
  \pg{173--195}.

\bibitem[Jordan {\em et~al.\/}(2017)Jordan, Zhang, Lehnasch \&
  Cavalieri]{jordan2017modal}
{\sc \au{Jordan, P. }, \au{Zhang, M. }, \au{Lehnasch, G. } \& \au{Cavalieri,
  A.~V. }} \yr{2017} Modal and non-modal linear wavepacket dynamics in
  turbulent jets.  \bt{In {\em 23rd AIAA/CEAS Aeroacoustics Conference\/}},
  \pg{p. 3379}.

\bibitem[Kaiser {\em et~al.\/}(2019)Kaiser, Lesshafft \&
  Oberleithner]{kaiser2019prediction}
{\sc \au{Kaiser, T.~L. }, \au{Lesshafft, L. } \& \au{Oberleithner, K. }}
  \yr{2019}  \at{Prediction of the flow response of a turbulent flame to
  acoustic pertubations based on mean flow resolvent analysis} .

\bibitem[Khorrami {\em et~al.\/}(1989)Khorrami, Malik \&
  Ash]{khorrami1989application}
{\sc \au{Khorrami, M.~R. }, \au{Malik, M.~R. } \& \au{Ash, R.~L. }} \yr{1989}
  \at{Application of spectral collocation techniques to the stability of
  swirling flows}.  \jt{Journal of Computational Physics}  \bvol{81}~(1),
  \pg{206--229}.

\bibitem[Kovesi(2015)]{kovesi2015good}
{\sc \au{Kovesi, P. }} \yr{2015}  \at{Good colour maps: How to design them}.
  \jt{arXiv preprint arXiv:1509.03700} .

\bibitem[Lesshafft \& Huerre(2007)]{lesshafft2007linear}
{\sc \au{Lesshafft, L. } \& \au{Huerre, P. }} \yr{2007}  \at{Linear impulse
  response in hot round jets}.  \jt{Physics of Fluids}  \bvol{19}~(2),
  \pg{024102}.

\bibitem[Lesshafft {\em et~al.\/}(2019)Lesshafft, Semeraro, Jaunet, Cavalieri
  \& Jordan]{lesshafft2019resolvent}
{\sc \au{Lesshafft, L. }, \au{Semeraro, O. }, \au{Jaunet, V. }, \au{Cavalieri,
  A.~V. } \& \au{Jordan, P. }} \yr{2019}  \at{Resolvent-based modeling of
  coherent wave packets in a turbulent jet}.  \jt{Physical Review Fluids}
  \bvol{4}~(6),  \pg{063901}.

\bibitem[Li \& Malik(1997)]{li1997spectral}
{\sc \au{Li, F. } \& \au{Malik, M.~R. }} \yr{1997}  \at{Spectral analysis of
  parabolized stability equations}.  \jt{Computers \& fluids}  \bvol{26}~(3),
  \pg{279--297}.

\bibitem[Lumley(1967)]{lumley1967structure}
{\sc \au{Lumley, J.~L. }} \yr{1967}  \at{The structure of inhomogeneous
  turbulent flows}.  \jt{Atmospheric turbulence and radio wave propagation} .

\bibitem[Maestrello \& McDaid(1971)]{maestrello1971acoustic}
{\sc \au{Maestrello, L. } \& \au{McDaid, E. }} \yr{1971}  \at{Acoustic
  characteristics of a high-subsonic jet}.  \jt{AIAA Journal}  \bvol{9}~(6),
  \pg{1058--1066}.

\bibitem[Martini {\em et~al.\/}(2020)Martini, Rodr{\'\i}guez, Towne \&
  Cavalieri]{martini2020efficient}
{\sc \au{Martini, E. }, \au{Rodr{\'\i}guez, D. }, \au{Towne, A. } \&
  \au{Cavalieri, A.~V. }} \yr{2020}  \at{Efficient computation of global
  resolvent modes}.  \jt{arXiv preprint arXiv:2008.10904} .

\bibitem[McKeon \& Sharma(2010)]{mckeon2010critical}
{\sc \au{McKeon, B. } \& \au{Sharma, A. }} \yr{2010}  \at{A critical-layer
  framework for turbulent pipe flow}.  \jt{Journal of Fluid Mechanics}
  \bvol{658},  \pg{336--382}.

\bibitem[Michalke(1971)]{michalke1971instabilitaet}
{\sc \au{Michalke, A. }} \yr{1971}  \bt{Instabilitaet eines kompressiblen
  runden freistrahis unter beruecksichtigung des einflusses der
  strahigrenzschichtdicke (instability of a compressible circular jet
  considering the influence of the thickness of the jet boundary layer)}. {\em
  Tech. Rep.\/}.  \org{DEUTSCHE FORSCHUNGS-UND VERSUCHSANSTALT FUER LUFT-UND
  RAUMFAHRT EV BERLIN~…}.

\bibitem[Michalke(1984)]{michalke1984survey}
{\sc \au{Michalke, A. }} \yr{1984}  \at{Survey on jet instability theory}.
  \jt{Progress in Aerospace Sciences}  \bvol{21},  \pg{159--199}.

\bibitem[Mollo-Christensen(1967)]{mollo1967}
{\sc \au{Mollo-Christensen, E. }} \yr{1967}  \at{{Jet Noise and Shear Flow
  Instability Seen From an Experimenter’s Viewpoint}}.  \jt{Journal of
  Applied Mechanics}  \bvol{34}~(1),  \pg{1--7}.

\bibitem[Monkewitz {\em et~al.\/}(2007)Monkewitz, Chauhan \&
  Nagib]{monkewitz2007self}
{\sc \au{Monkewitz, P.~A. }, \au{Chauhan, K.~A. } \& \au{Nagib, H.~M. }}
  \yr{2007}  \at{Self-consistent high-reynolds-number asymptotics for
  zero-pressure-gradient turbulent boundary layers}.  \jt{Physics of Fluids}
  \bvol{19}~(11),  \pg{115101}.

\bibitem[Moore(1977)]{moore1977role}
{\sc \au{Moore, C. }} \yr{1977}  \at{The role of shear-layer instability waves
  in jet exhaust noise}.  \jt{Journal of Fluid Mechanics}  \bvol{80}~(2),
  \pg{321--367}.

\bibitem[Morra {\em et~al.\/}(2020)Morra, Nogueira, Cavalieri \&
  Henningson]{morra2020colour}
{\sc \au{Morra, P. }, \au{Nogueira, P.~A. }, \au{Cavalieri, A.~V. } \&
  \au{Henningson, D.~S. }} \yr{2020}  \at{The colour of forcing statistics in
  resolvent analyses of turbulent channel flows}.  \jt{arXiv preprint
  arXiv:2004.01565} .

\bibitem[Muralidhar {\em et~al.\/}(2019)Muralidhar, Podvin, Mathelin \&
  Fraigneau]{muralidhar2019spatio}
{\sc \au{Muralidhar, S.~D. }, \au{Podvin, B. }, \au{Mathelin, L. } \&
  \au{Fraigneau, Y. }} \yr{2019}  \at{Spatio-temporal proper orthogonal
  decomposition of turbulent channel flow}.  \jt{Journal of Fluid Mechanics}
  \bvol{864},  \pg{614--639}.

\bibitem[Nogueira {\em et~al.\/}(2020{\natexlab{{\em a\/}}})Nogueira,
  Cavalieri, Hanifi \& Henningson]{nogueira2020resolvent}
{\sc \au{Nogueira, P.~A. }, \au{Cavalieri, A.~V. }, \au{Hanifi, A. } \&
  \au{Henningson, D.~S. }} \yr{2020{\natexlab{{\em a\/}}}}  \at{Resolvent
  analysis in unbounded flows: role of free-stream modes}.  \jt{Theoretical and
  Computational Fluid Dynamics}  \pg{pp. 1--14}.

\bibitem[Nogueira {\em et~al.\/}(2019)Nogueira, Cavalieri, Jordan \&
  Jaunet]{nogueira2019large}
{\sc \au{Nogueira, P.~A. }, \au{Cavalieri, A.~V. }, \au{Jordan, P. } \&
  \au{Jaunet, V. }} \yr{2019}  \at{Large-scale streaky structures in turbulent
  jets}.  \jt{Journal of Fluid Mechanics}  \bvol{873},  \pg{211--237}.

\bibitem[Nogueira {\em et~al.\/}(2020{\natexlab{{\em b\/}}})Nogueira, Morra,
  Martini, Cavalieri \& Henningson]{nogueira2020forcing}
{\sc \au{Nogueira, P.~A. }, \au{Morra, P. }, \au{Martini, E. }, \au{Cavalieri,
  A.~V. } \& \au{Henningson, D.~S. }} \yr{2020{\natexlab{{\em b\/}}}}
  \at{Forcing statistics in resolvent analysis: application in minimal
  turbulent couette flow}.  \jt{arXiv preprint arXiv:2001.02576} .

\bibitem[Passaggia {\em et~al.\/}(2009)Passaggia, Ehrenstein \&
  Gallaire]{passaggia2009control}
{\sc \au{Passaggia, P.-Y. }, \au{Ehrenstein, U. } \& \au{Gallaire, F. }}
  \yr{2009} Control of a separated boundary layer: model reduction using global
  modes revisited.  \bt{In {\em 4th Global Instability and Control
  Symposium\/}},  \pg{pp. CD--ROM}.

\bibitem[Pickering {\em et~al.\/}(2020)Pickering, Rigas, Nogueira, Cavalieri,
  Schmidt \& Colonius]{pickering2020lift}
{\sc \au{Pickering, E. }, \au{Rigas, G. }, \au{Nogueira, P. A.~S. },
  \au{Cavalieri, A. V.~G. }, \au{Schmidt, O.~T. } \& \au{Colonius, T. }}
  \yr{2020}  \at{Lift-up, kelvin-helmholtz and orr mechanisms in turbulent
  jets}.  \jt{Journal of Fluid Mechanics}  \bvol{896},  \pg{A2}.

\bibitem[Rodr{\'\i}guez {\em et~al.\/}(2015)Rodr{\'\i}guez, Cavalieri, Colonius
  \& Jordan]{rodriguez2015study}
{\sc \au{Rodr{\'\i}guez, D. }, \au{Cavalieri, A.~V. }, \au{Colonius, T. } \&
  \au{Jordan, P. }} \yr{2015}  \at{A study of linear wavepacket models for
  subsonic turbulent jets using local eigenmode decomposition of piv data}.
  \jt{European Journal of Mechanics-B/Fluids}  \bvol{49},  \pg{308--321}.

\bibitem[Salwen \& Grosch(1981)]{salwen1981continuous}
{\sc \au{Salwen, H. } \& \au{Grosch, C.~E. }} \yr{1981}  \at{The continuous
  spectrum of the orr-sommerfeld equation. part 2. eigenfunction expansions}.
  \jt{Journal of Fluid Mechanics}  \bvol{104},  \pg{445--465}.

\bibitem[Sano {\em et~al.\/}(2019)Sano, Abreu, Cavalieri \&
  Wolf]{sano2019trailing}
{\sc \au{Sano, A. }, \au{Abreu, L.~I. }, \au{Cavalieri, A.~V. } \& \au{Wolf,
  W.~R. }} \yr{2019}  \at{Trailing-edge noise from the scattering of
  spanwise-coherent structures}.  \jt{Physical Review Fluids}  \bvol{4}~(9),
  \pg{094602}.

\bibitem[Schlatter \& {\"O}rl{\"u}(2010)]{schlatter2010assessment}
{\sc \au{Schlatter, P. } \& \au{{\"O}rl{\"u}, R. }} \yr{2010}  \at{Assessment
  of direct numerical simulation data of turbulent boundary layers}.
  \jt{Journal of Fluid Mechanics}  \bvol{659},  \pg{116--126}.

\bibitem[Schlichting \& Gersten(2016)]{schlichting2016boundary}
{\sc \au{Schlichting, H. } \& \au{Gersten, K. }} \yr{2016} {\em Boundary-layer
  theory\/}.  \publ{Springer}.

\bibitem[Schmidt {\em et~al.\/}(2018)Schmidt, Towne, Rigas, Colonius \&
  Br{\`e}s]{schmidt2018spectral}
{\sc \au{Schmidt, O.~T. }, \au{Towne, A. }, \au{Rigas, G. }, \au{Colonius, T. }
  \& \au{Br{\`e}s, G.~A. }} \yr{2018}  \at{Spectral analysis of jet
  turbulence}.  \jt{Journal of Fluid Mechanics}  \bvol{855},  \pg{953--982}.

\bibitem[Semeraro {\em et~al.\/}(2016)Semeraro, Lesshafft, Jaunet \&
  Jordan]{semeraro2016modeling}
{\sc \au{Semeraro, O. }, \au{Lesshafft, L. }, \au{Jaunet, V. } \& \au{Jordan,
  P. }} \yr{2016}  \at{Modeling of coherent structures in a turbulent jet as
  global linear instability wavepackets: Theory and experiment}.
  \jt{International Journal of Heat and Fluid Flow}  \bvol{62},  \pg{24--32}.

\bibitem[Sharma \& McKeon(2013)]{sharma2013coherent}
{\sc \au{Sharma, A.~S. } \& \au{McKeon, B.~J. }} \yr{2013}  \at{On coherent
  structure in wall turbulence}.  \jt{Journal of Fluid Mechanics}  \bvol{728},
  \pg{196--238}.

\bibitem[Smits {\em et~al.\/}(2011)Smits, McKeon \& Marusic]{smits2011high}
{\sc \au{Smits, A.~J. }, \au{McKeon, B.~J. } \& \au{Marusic, I. }} \yr{2011}
  \at{High--reynolds number wall turbulence}.  \jt{Annual Review of Fluid
  Mechanics}  \bvol{43}.

\bibitem[Suzuki \& Colonius(2006)]{suzuki2006instability}
{\sc \au{Suzuki, T. } \& \au{Colonius, T. }} \yr{2006}  \at{Instability waves
  in a subsonic round jet detected using a near-field phased microphone array}.
   \jt{Journal of Fluid Mechanics}  \bvol{565},  \pg{197--226}.

\bibitem[Tissot {\em et~al.\/}(2017{\natexlab{{\em a\/}}})Tissot, Laj{\'u}s~Jr,
  Cavalieri \& Jordan]{tissot2017wave}
{\sc \au{Tissot, G. }, \au{Laj{\'u}s~Jr, F.~C. }, \au{Cavalieri, A.~V. } \&
  \au{Jordan, P. }} \yr{2017{\natexlab{{\em a\/}}}}  \at{Wave packets and orr
  mechanism in turbulent jets}.  \jt{Physical Review Fluids}  \bvol{2}~(9),
  \pg{093901}.

\bibitem[Tissot {\em et~al.\/}(2017{\natexlab{{\em b\/}}})Tissot, Zhang,
  Laj{\'u}s, Cavalieri \& Jordan]{tissot2017sensitivity}
{\sc \au{Tissot, G. }, \au{Zhang, M. }, \au{Laj{\'u}s, F.~C. }, \au{Cavalieri,
  A.~V. } \& \au{Jordan, P. }} \yr{2017{\natexlab{{\em b\/}}}}  \at{Sensitivity
  of wavepackets in jets to nonlinear effects: the role of the critical layer}.
   \jt{Journal of Fluid Mechanics}  \bvol{811},  \pg{95--137}.

\bibitem[Towne {\em et~al.\/}(2018)Towne, Schmidt \&
  Colonius]{towne2018spectral}
{\sc \au{Towne, A. }, \au{Schmidt, O.~T. } \& \au{Colonius, T. }} \yr{2018}
  \at{Spectral proper orthogonal decomposition and its relationship to dynamic
  mode decomposition and resolvent analysis}.  \jt{Journal of Fluid Mechanics}
  \bvol{847},  \pg{821--867}.

\bibitem[Tumin(1996)]{tumin1996receptivity}
{\sc \au{Tumin, A. }} \yr{1996}  \at{Receptivity of pipe poiseuille flow}.
  \jt{Journal of Fluid Mechanics}  \bvol{315},  \pg{119--137}.

\bibitem[Tumin \& Fedorov(1983)]{tumin1983spatial}
{\sc \au{Tumin, A. } \& \au{Fedorov, A. }} \yr{1983}  \at{Spatial growth of
  disturbances in a compressible boundary layer}.  \jt{Journal of Applied
  Mechanics and Technical Physics}  \bvol{24}~(4),  \pg{548--554}.

\bibitem[Viswanathan \& Clark(2004)]{viswanathan2004effect}
{\sc \au{Viswanathan, K. } \& \au{Clark, L. }} \yr{2004}  \at{Effect of nozzle
  internal contour on jet aeroacoustics}.  \jt{International Journal of
  Aeroacoustics}  \bvol{3}~(2),  \pg{103--135}.

\bibitem[Zaman(1985)]{zaman1985effect}
{\sc \au{Zaman, K. }} \yr{1985}  \at{Effect of initial condition on subsonic
  jet noise}.  \jt{AIAA journal}  \bvol{23}~(9),  \pg{1370--1373}.

\bibitem[Zaman(2012)]{zaman2012effect}
{\sc \au{Zaman, K. }} \yr{2012}  \at{Effect of initial boundary-layer state on
  subsonic jet noise}.  \jt{AIAA journal}  \bvol{50}~(8),  \pg{1784--1795}.

\bibitem[Zaman \& Hussain(1981)]{zaman1981turbulence}
{\sc \au{Zaman, K. } \& \au{Hussain, A. }} \yr{1981}  \at{Turbulence
  suppression in free shear flows by controlled excitation}.  \jt{Journal of
  Fluid Mechanics}  \bvol{103},  \pg{133--159}.

\end{thebibliography}
\bibliographystyle{jfm}
\end{document}